\documentclass[aps,preprint,nofootinbib,floatfix]{revtex4}
\usepackage{graphicx,color}
\usepackage[latin1]{inputenc}
\usepackage{amsmath,amssymb}
\usepackage{hyperref}
\usepackage{epstopdf}
\usepackage{slashed}

\newcommand{\lsim}{\mathrel{\mathop{\kern 0pt \rlap
  {\raise.2ex\hbox{$<$}}}
  \lower.9ex\hbox{\kern-.190em $\sim$}}}
\newcommand{\gsim}{\mathrel{\mathop{\kern 0pt \rlap
  {\raise.2ex\hbox{$>$}}}
  \lower.9ex\hbox{\kern-.190em $\sim$}}}

\newcommand{\sigmav}{\ensuremath{\langle\sigma v\rangle}}

\newcommand{\hinv}{\ensuremath{{\rm BR}(h\to \texttt{inv.})}}

\newcommand{\sigsip}{\ensuremath{\sigma^{\rm{SI}}_p}}

\newcommand{\kev}{\ensuremath{\,\mathrm{keV}}}
\newcommand{\mev}{\ensuremath{\,\mathrm{MeV}}}
\newcommand{\gev}{\ensuremath{\,\mathrm{GeV}}}

\def  \bcen   {\begin{center}}
\def  \ecen   {\end{center}}
\def  \beq    {\begin{equation}}
\def  \eeq    {\end{equation}}
\def  \beqa   {\begin{eqnarray}}
\def  \eeqa   {\end{eqnarray}}

\def\bea{\begin{eqnarray}}
\def\eea{\end{eqnarray}}

\begin{document}
\title{Confronting dark matter co-annihilation of Inert two Higgs Doublet Model with a compressed mass spectrum}
\author{Chih-Ting Lu$^1$\footnote{timluyu@kias.re.kr}}
\author{Van Que Tran$^2$\footnote{vqtran@nju.edu.cn}}
\author{Yue-Lin Sming Tsai$^{3,4}$\footnote{smingtsai@gate.sinica.edu.tw}}
\affiliation{
$^1$School of Physics, KIAS, Seoul 130-722, Republic of Korea \\
$^2$School of Physics, Nanjing University, Nanjing 210093, China \\
$^3$Institute of Physics, Academia Sinica, Nangang, Taipei 11529, Taiwan \\
$^4$~Key Laboratory of Dark Matter and Space Astronomy,  
Purple Mountain Observatory, Chinese Academy of Sciences, Nanjing 210008, China
}

\date{\today}

\begin{abstract}
We perform a comprehensive analysis 
for the light scalar dark matter (DM)
in the Inert two Higgs doublet model (i2HDM) with 
compressed mass spectra,   
small mass splittings among three $\mathbb{Z}_2$ odd particles---scalar $S$, pseudo-scalar $A$, 
and charged Higgs $H^\pm$. 
In such a case,
the co-annihilation processes 
play a significant role to reduce DM relic density. 
As long as a co-annihilation governs the total interaction rate in the early universe, 
a small annihilation rate is expected to reach a correct DM relic density and its coupling 
$\lambda_S$ between DM pair and Higgs boson shall be tiny.
Consequently, a negligible DM-nucleon elastic scattering cross section is predicted at the tree-level.  
In this work, we include the one-loop quantum corrections 
of the DM-nucleon elastic scattering cross section. 
We found that the quartic self-coupling $\lambda_2$ between $\mathbb{Z}_2$ odd particles indeed contributes 
the one-loop quantum correction 
and behaves non-trivially for the co-annihilation scenario.  
Interestingly, the parameter space, which is allowed by the current constraints 
considered in this study,
can predict the DM mass 
and annihilation cross section at the present 
compatible with the AMS-02 antiproton excess.
The parameter space can be further probed  
at the future high luminosity LHC. 


\end{abstract}

\maketitle

\section{Introduction \label{section:1}}

Roughly a quarter of the Universe is made of Dark Matter (DM),
but many experimental results also reveal that DM is weakly 
or even not interacting with the Standard Model (SM) sector except for the gravitational force.  
To understand the particle nature of DM, 
several detection methods have been developed in the past decades 
such as DM direct detection (DD), indirect detection (ID) and colliders. 
Although some of DD and ID analyses have reported anomalies
\cite{Bernabei:2010mq,Aalseth:2010vx,TheFermi-LAT:2017vmf,Cui:2016ppb,Cuoco:2016eej}, 
DM signal is still absent in the LHC searches~\cite{Felcini:2018osp} 
so that the DM properties (e.g., spin, mass, and couplings) 
are still not able to be determined. 
By considering interactions between DM and SM particles 
within the detector energy threshold, 
two possibilities arise from null signal detection. 
The first possibility is due to the couplings between DM and SM particles  
are too weak to be detected under the current sensitivities. 
Hence, an upgrade design of instrument and longer time of exposure may be needed 
in order to catch the DM signal~\cite{Richard:2014vfa,Han:2018wus}. 
Nevertheless, if a very tiny coupling between DM and SM particles 
would be favored by future measurements, 
the current search strategies for the weakly interacting massive particles (WIMPs) 
are still hard to work~\cite{Arcadi:2017kky}.
The second possibility comes from the compressed mass spectrum models 
in which the next lightest dark particle and DM 
have a small mass splitting~\cite{Griest:1990kh}.
The signals from such compressed mass spectra are usually predicted with soft objects 
and then vetoed or polluted 
from SM backgrounds at colliders~\cite{Harland-Lang:2018hmi}.
Due to the small mass splitting, 
the next lightest dark particle can be long lived.
Interestingly, the interaction couplings 
between dark sector and SM fields may be not suppressed 
for this possibility 
and they can be testable in the DD or ID searches~\cite{Okada:2019sbb}.

The inert two Higgs doublet model (i2HDM)
\cite{Deshpande:1977rw,Ma:2006km, Barbieri:2006dq, LopezHonorez:2006gr} 
is a simplest spin-zero DM model within the framework of two Higgs doublets,  
and it can naturally realize the compressed mass spectrum~\cite{Blinov:2015qva}. 
There are three $\mathbb{Z}_2$-odd scalar bosons: 
scalar $S$, pseudo-scalar $A$, and charged Higgs $H^\pm$.
Either $S$ or $A$ can play the role of a DM candidate, 
but it is difficult to distinguish one from the other in the 
phenomenological point of view~\cite{Arhrib:2013ela}. 
In this paper, we only discuss the scalar $S$ playing the role of the DM candidate.   
The discrete $\mathbb{Z}_2$ symmetry in i2HDM can be taken as an accident symmetry 
after the symmetry breaking of a larger continuous symmetry. 
These residual approximated symmetries of the scalar potential 
can force the mass spectrum of these exotic scalars to be compressed. 
Given the fact that the SM Higgs doublet $H_1$ is $\mathbb{Z}_2$-even 
and the second doublet $H_2$ is $\mathbb{Z}_2$-odd, 
one can have $ m_S = m_A $ at the leading order as long as 
the term $\lambda_5 \left[ (H_1^\dagger H_2)^2 +h.c.\right]$ 
in the scalar potential is vanished or omitted \cite{Barbieri:2006dq}. 
Similarly, if further removing the term $\lambda_4 (H_2^\dagger H_1)(H_2^\dagger H_1)$ 
in the scalar potential, 
all three of $\mathbb{Z}_2$-odd scalars are degenerated at the leading order~\cite{Gerard:2007kn}. 
Hence, the compressed mass spectrum can be naturally realized in the i2HDM.

As the thermal DM scenario, a compressed mass spectrum usually results a sufficient co-annihilation in the early universe 
because the lightest and the next-lightest $\mathbb{Z}_2$-odd particles are with considerable number density before freeze-out~\cite{Griest:1990kh}. 
Requiring the correct relic density to be in agreement with the PLANCK collaboration~\cite{Aghanim:2018eyx}, 
the sum of WIMPs annihilation and co-annihilation rates shall be within a certain range. 
If the co-annihilation rate overwhelms the total interaction and the annihilation is inactive, 
the coupling $\lambda_S$ between DM pair and SM Higgs boson will be 
very small which may lead to undetectable signals at the current DD and collider searches. 
In particular, due to the smallness of $\lambda_S$, the DM-nucleon elastic cross section 
is suppressed at tree level so that some regions of parameter space cannot be probed by the current DDs. 
As reported in Ref.~\cite{Abe:2015rja}, 
the elastic scattering cross section can be enhanced at the loop level which non-trivially
depends on the size of quartic self-coupling $\lambda_2$ between $\mathbb{Z}_2$ odd bosons and the mass splittings, 
$\Delta^0=m_A-m_S$ and $\Delta^\pm=m_{H^\pm}-m_S$. 
Comparing with Ref.~\cite{Abe:2015rja} in which only the Higgs resonance regions are discussed, 
we further investigate the impact of the loop corrections on the co-annihilation scenario in this work.

In this paper, a global analysis of the compressed mass spectrum scenario 
within the framework of the i2HDM is performed under the combined constraints 
from theoretical conditions, collider searches, relic density, XENON1T, and Fermi dSphs gamma ray data.  
In order to highlight the role of compressed mass spectrum in the early universe, 
we further divide the allowed region into two groups: the co-annihilation and mixed scenarios.  
This classification is based on the DM relic density reduction before freeze-out 
which is mainly governed by the annihilation or co-annihilation processes. 
The leading order (LO) and next-leading order (NLO) contributions to 
the DM-nucleon elastic scattering cross section are also computed and compared 
in the context of these two scenarios. 
In some regions of the model parameter space, particularly for the co-annihilation scenario, 
the DM spin-independent cross section at the NLO can be significantly enhanced and 
hence probed by the present XENON1T result.   
Including this NLO enhancement,
one can pin down the parameter space of $m_S$, $\lambda_2$, $\Delta^0$ and $\Delta^\pm$.  
The surviving parameter space is also compatible with AMS-02 antiproton anomaly and 
can be further probed at the future DD and the high luminosity LHC (HL-LHC) 
searches for the compressed mass spectrum.

The remaining sections of the paper are organized as follows. In Sec.~\ref{section:2}, 
we briefly revisit the i2HDM model, the corrections of mass spectrum beyond the tree-level, 
and the decay width of heavier $\mathbb{Z}_2$-odd particles. 
In Sec.~\ref{sec:constraints}, we consider both the theoretical and experimental constraints
used in our likelihood functions. 
In Sec.~\ref{section:result}, we present our numerical analysis and the $2\sigma$ allowed regions with 
and without the loop corrections to the DM-nucleon scattering calculation. 
Finally, we conclude in Sec.~\ref{sec:conclusion}. 
 
\section{Inert two Higgs Doublet Model~\label{section:2}}

In this section, we first review the structure of i2HDM and its model parameters.
We then discuss the possible one-loop contributions  
of $\Delta^0$ and $\Delta^\pm$, 
including renormalization group equations (RGEs) and electroweak symmetry breaking (EWSB). 
Finally, the decay widths of $A$ and $H^{\pm}$ in the case of compressed mass spectrum are given in Sec.~\ref{section:2c}.


\subsection{Parameterization of the i2HDM scalar potential\label{section:2a}}

The i2HDM~\cite{Deshpande:1977rw} is the simplest version of DM model within 
the two Higgs doublets framework. 
Compared with the single scalar doublet in the SM,
the i2HDM has two scalar doublets $H_1$ and $H_2$ under 
a discrete $\mathbb{Z}_2$ symmetry, $H_1\rightarrow H_1$ and $H_2\rightarrow -H_2$ 
which is introduced to maintain the stability of DM.
The $\mathbb{Z}_2$ symmetry cannot be spontaneously broken 
so that $H_2$ never develops a vacuum expectation value (VEV).
These two doublets can be given as   
\beq
H_1 =
   \left( \begin{array}{c}  G^+ \\ \frac{1}{\sqrt{2}} \left( v + 
h + i G^0 \right)  \end{array}
     \right), ~~{\rm and}~~
H_2 =    \left( \begin{array}{c}  H^+ \\ 
\frac{1}{\sqrt 2} (S  + i A)  \end{array}  \right).  
\eeq
Here, $G^\pm$ and $G^0$ are charged and neutral Goldstone bosons respectively.
The symmetry breaking pattern for the doublets are 
$\langle H_1^T \rangle = \left( 0 , \; v/\sqrt{2} \right)$ 
and $\langle H_2^T \rangle = \left( 0 , \; 0 \right)$, 
where $v\approx 246\gev$.
In the end, we have five physical mass eigenstates: 
two CP-even neutral scalar $h$ and $S$, 
one CP-odd neutral scalar $A$, and a pair of charged scalars $H^{\pm}$.

Before going to the detailed calculation,  
let us briefly recap the main features of the i2HDM. 
First, $\mathbb{Z}_2$-odd particles $S$, $A$ and $H^{\pm}$ 
are not directly coupled to SM fermions 
while $\mathbb{Z}_2$-even Higgs $h$ 
is identified as 
the SM Higgs with mass $\sim 125\gev$. 
Second, owing to the exact $\mathbb{Z}_2$ symmetry, 
there is no tree-level flavor changing neutral current. 
Finally, the DM candidate can be either $S$ or $A$ 
depending on their masses, 
but it is hard to phenomenologically distinguish one from the other~\cite{Arhrib:2013ela}. 
Here, we restrict ourselves to focus on the CP-even scalar $S$ 
as the DM candidate rather than the CP-odd pseudo scalar $A$.

Unlike the general two Higgs doublet model, 
since the mixing term $-\mu_{12}^2 (H_1^\dagger H_2 + {\rm h.c.})$ 
is forbidden by the exact $\mathbb{Z}_2$ symmetry,
the scalar potential of i2HDM has a simpler form, 
\begin{eqnarray}
\label{potential}
V &=& \mu_1^2 |H_1|^2 + \mu_2^2 |H_2|^2 + \lambda_1 |H_1|^4
+ \lambda_2 |H_2|^4 +  \lambda_3 |H_1|^2 |H_2|^2 + \lambda_4
|H_1^\dagger H_2|^2 \nonumber \\
&& \;\;\;\; + \; \frac{\lambda_5}{2} \left\{ (H_1^\dagger H_2)^2 + {\rm h.c.} \right\} \;\; .
\end{eqnarray}
After the electroweak symmetry breaking, there remains eight real parameters: 
five $\lambda$s, $\mu_1$, $\mu_2$ and the VEV $v$ for the scalar potential.  
Because two parameters the VEV $v$ and Higgs mass can be fixed by the experimental observations and  
one parameter can be eliminated by the Higgs potential minimum condition, 
only five real parameters ($\mu_2^2$, $\lambda_2$, $\lambda_3$, $\lambda_4$ and $\lambda_5$) 
are inputs of the model.  
Note that the quartic coupling $\lambda_2$ is only involved in the four-points interaction of 
$\mathbb{Z}_2$-odd scalar bosons ($|H_2|^4$), which is a phenomenologically invisible interaction at the tree-level. 
Nevertheless, the role of $\lambda_2$ 
is important to the calculations of the DM-nuclei elastic scattering cross section 
at the one-loop level~\cite{Abe:2015rja}.

Conventionally, it is more intuitive to adopt the physical mass basis as inputs 
\begin{eqnarray}
\label{mh-lambda1}
&& m_h^2=-2 \mu_1^2=2 \lambda_1 v^2, \nonumber\\
&& m_S^2=\mu_2^2+\frac{1}{2} (\lambda_3 +\lambda_4 + \lambda_5) v^2=
\mu_2^2+ \lambda_S v^2, \nonumber\\
&& m_A^2=\mu_2^2+\frac{1}{2} (\lambda_3 +\lambda_4 - \lambda_5) v^2 = 
\mu_2^2+ \lambda_A v^2, \nonumber\\
&& m_{H^\pm}^2= \mu_2^2+ \frac{1}{2} \lambda_3  v^2,  
\end{eqnarray}
where we denote 
\beq
\lambda_{S}=\frac{1}{2}(\lambda_3 +\lambda_4 + \lambda_5),~~{\rm and}~~ 
\lambda_A = \lambda_S - \lambda_5 = \lambda_S + \frac{m_A^2-m_S^2}{v^2}.
\eeq
Assuming $m_S<m_A$, we can see that 
$\lambda_S$ is always smaller than $\lambda_A$ at the tree-level parameterization.  
Reversely, the quartic couplings $\lambda_{i=1,3,4,5}$ in terms of these 4 physical scalar masses 
and $\mu^2_2$ are given by   

\beqa
\lambda_1 &=& \frac{m_h^2}{2 v^2}\quad , \quad 
\lambda_3 = \frac{2}{v^2}\left(m_{H^\pm}^2 - \mu_2^2\right), \nonumber\\
\lambda_4 &=& \frac{\left(m_S^2 + m_A^2 - 2 m_{H^\pm}^2\right)}{v^2} \quad ,
\quad 
\lambda_5 = \frac{\left(m_S^2 - m_A^2\right)}{v^2} \;\; .
\label{lambds}
\eeqa
For the scenario with the compressed mass spectra, 
the mass splitting parameters 
$\Delta^0 =m_A -m_S$ and $\Delta^{\pm} =m_{H^{\pm}} -m_S$ instead of $m_A$ and $m_{H^\pm}$ 
are more useful. 
Hence, our input parameters are 
\beq
\{m_S,\Delta^0, \Delta^{\pm}, \lambda_2 , \lambda_S \}.
\label{eq:input_par_3}
\eeq


\subsection{Scalar mass splittings beyond the tree level\label{section:2b}}

Considering a compressed mass spectrum in the i2HDM, namely small $\Delta^0$ and $\Delta^{\pm}$, 
the couplings $\lambda_4$ and $\lambda_5$ are naturally small as shown in Eq.~(\ref{lambds}).  
However, possible modifications to $\lambda_4$ and $\lambda_5$ from renormalization group equations (RGEs)
beyond the tree-level may not be ignored. 
The one-loop RGEs of $\lambda_4$ and $\lambda_5$ are represented as~\cite{Goudelis:2013uca, Blinov:2015qva}
\begin{eqnarray}
(4\pi)^2\frac{d\lambda_4}{dt} = && -3\lambda_4(3g^2+g^{\prime 2})
+ 4\lambda_4(\lambda_1+\lambda_2+2\lambda_3+\lambda_4)\nonumber\\
&& +2\lambda_4(3y_t^2+3y_b^2+y_{\tau}^2)
+3g^2g^{\prime 2}+8\lambda_5^2,\label{eq:l4run} \\
(4\pi)^2\frac{d\lambda_5}{dt} = && -3\lambda_5(3g^2+g^{\prime 2})
+ 4\lambda_5(\lambda_1+\lambda_2+2\lambda_3+3\lambda_4)\nonumber \\
&& +2\lambda_5(3y_t^2+3y_b^2+y_{\tau}^2).
\label{eq:l5run}
\end{eqnarray}
where $t=\ln(\mu/\mu_0)$ with renormalization scale $\mu$   
divided by the electroweak scale $\mu_0 =100\gev$.
Since all terms in the right hand side of Eq.~\eqref{eq:l5run} are proportional to $\lambda_5$, 
once we set $\lambda_5 =0$ at any reference scale, its value does not change in the one-loop RGEs. 
Still, there is one term proportional to $g^2g^{\prime 2}$ 
in the right hand side of Eq.~\eqref{eq:l4run};  
Even if we set $\lambda_4=\lambda_5 =0$ at a specific reference scale, 
the value of $\lambda_4$ can be modified by the one-loop RGEs. 
Therefore, unlike the neutral mass splitting $\Delta^{0}$, 
the charged mass splitting $\Delta^{\pm}$ cannot be extremely small. 
More details can be found in Ref.~\cite{Blinov:2015qva}.

Additionally, there is a finite contribution to $\Delta^{\pm}$ from EWSB at one-loop level~\cite{Cirelli:2005uq},  
and it is given by 
\beq
\delta\Delta^{\pm} ~=~ 
\frac{g^2 \sin^2 \theta_W}{16\pi^2}m_{H^{\pm}}\,f(m_Z/m_{H^{\pm}}), 
\eeq
where $f(x)$ is defined as 
\beq
f(x) = -\frac{x}{4}\left(
2x^3\ln x 
+(x^2-4)^{3/2}\ln\left[(x^2-2-x\sqrt{x^2-4})/2\right]
\right).
\eeq
This extra mass splitting can be at most about $\mathcal{O}(100)\mev$~\cite{Cirelli:2005uq, Blinov:2015qva} and it is usually smaller than the one from RGEs.

As aforementioned, 
$\Delta^{0}$ can be very small if the discrete $\mathbb{Z}_2$ symmetry on $H_2$ coming from global $U(1)$ symmetry~\cite{Barbieri:2006dq}. 
The possible one-loop contributions for $\Delta^{0}$ can be neglected once $\lambda_5\sim 0$ at any reference scale. 
On the other hand, even if we set $\lambda_4 =\lambda_5 =0$ at a specific reference scale 
based on global $SU(2)$ symmetry or custodial symmetry on $H_2$ in the i2HDM~\cite{Gerard:2007kn}, 
$\Delta^{\pm}$ may still have a correction of several hundred MeV from the one-loop contributions.
However, the effects from the loop corrections can be safely ignored in this analysis,  
as we require $\Delta^{\pm}\geq 1\gev$. 
Indeed, if one takes the one-loop corrections on $\Delta^0$ and $\Delta^\pm$ into account, 
the parameter space can only be slightly shifted but our result remains unchanged. 
Thus, we do not include these corrections in this analysis.


\subsection{Decay widths of $A$ and $H^{\pm}$ in compressed mass spectra\label{section:2c}}
In the compressed mass spectra of i2HDM, the dominant decay modes for $A$ and $H^{\pm}$ are $A\to SZ^*$ and $H^\pm\to SW^{\pm*}$, with off-shell $W/Z$ bosons.
After integrating out $W$ and $Z$ bosons, 
the decay widths for  $A\to Sf\bar{f}$ and $H^{\pm}\to Sf\bar{f}^\prime$ channels 
can be approximately given by
\beq
\Gamma(A\to Sf\bar{f}) ~=~ \frac{1}{120\pi^3}\frac{g^4}{m_W^4}(\Delta^{0})^5\,
\sum_iN_c^i\left[(a_V^i)^2+(a_A^i)^2\right]\times\Theta(\Delta^0-2m_i), 
\label{eq:gama}
\eeq
\beq
\Gamma(H^{\pm}\to Sf\bar{f}^\prime) ~=~ \frac{1}{120\pi^3}\frac{g^4}{m_W^4}(\Delta^{\pm})^5\,
\sum_{jk}N_c^{j}\left[|c_V^{\;jk}|^2+|c_A^{\;jk}|^2\right]
\times\Theta(\Delta^\pm-m_j-m_k), 
\label{eq:gamhp}
\eeq
where $N_c^{i(j)}$ is the color factor of the $i(j)$-th species
and $f$, $f^\prime$ are SM fermions. 
The step function $\Theta$ comes from the four-momentum conservation. 
The couplings $ a_V^i $ and $ a_A^i $ can be expressed as
\beq
a_V^i = \frac{1}{2}(T^3_i-2Q_is^2_W),~~~~~a_A^i = -\frac{1}{2}T^3_i \ ,
\eeq
where $i$ runs over all SM fermion species, $Q_i (T^3_i)$ 
is the charge (third component of isospin) 
for the $i$-th species, 
and $s_W$ stands for $\sin \theta_W$ with $\theta_W$ being the weak mixing angle. 
Similarly, the couplings $c_V^{jk}$ and $c_A^{jk}$ for lepton sectors can be represented as
\beq
c_V^{jk} ~=~ -c_A^{jk} =\frac{1}{2\sqrt{2}}\delta^{jk} \ ,
\eeq
and for quark sectors 
\beq
c_V^{jk} ~=~ -c_A^{jk} =\frac{1}{2\sqrt{2}}V^{jk}_{\rm CKM} \ ,
\eeq
where $j$ ($k$) runs over up-type (down-type) fermions
and $V_{\rm CKM}$ is the Cabbibo-Kobayashi-Maskawa matrix. 
We can apply the similar expression for the decay mode $H^{\pm}\to Af\bar{f}^\prime$.

The Eq.~\eqref{eq:gama} and Eq.~\eqref{eq:gamhp} show that 
the lifetimes of $A$ and $H^{\pm}$ are sensitive to $\Delta^0$ and $\Delta^{\pm}$, respectively.
For example, if $\Delta^0 < 2.6 $ GeV, 
the decay width $\Gamma(A\to Sf\bar{f}) < 2\times 10^{-10}$ GeV implies that the lifetime of $A$ 
is longer than the long-lived particle criterion at the LHC, $\sim 1~\mu {\rm m}/c$. 
In this analysis, nevertheless, both $\Delta^0$ and $\Delta^{\pm}$ are required to be larger
than$~5\gev$ due to the current constraints. Therefore, $A$ and $H^{\pm}$ cannot be long-lived particles.


\section{Constraints}
\label{sec:constraints}
In this section, we summarize the theoretical and experimental constraints used in our analysis. 
First, the theoretical constraints for the i2HDM Higgs potential 
such as the perturbativity, stability, and unitarity will be discussed. 
For the current experimental constraints, we will consider 
the collider, relic density, DM direct detection and DM indirect detection constraints.

\subsection{Theoretical constraints}
\label{section:theo}

Once the extra Higgs doublet has been introduced, the theoretical constraints of Higgs potential in i2HDM, 
such as the perturbativity, stability, and tree-level unitarity, have to be properly taken into account. 
As studied in the literature~\cite{Eriksson:2009ws}, these theoretical constraints are generically 
implemented in the Higgs basis $\lambda_{i}$ parameters. 
However, in this analysis, we use the physical mass basis as our inputs 
except for $\lambda_2$ and $\lambda_S=(\lambda_3+\lambda_4+\lambda_5)/2$. 
We employ the mass spectrum calculator \texttt{2HDMC}~\cite{Eriksson:2009ws} to make 
the conversion between these two bases and take care of the Higgs potential theoretical constraints.
We collect those parameter points which have passed the perturbativity, stability, and tree-level unitarity constraints.

\subsection{Collider Constraints}
 \label{section:3}
\subsubsection{\bf Electroweak precision tests} 
\label{sec:EWPT}

In the i2HDM, electroweak precision test (EWPT) is sensitive to the mass splitting among 
these $\mathbb{Z}_2$ odd scalar bosons \cite{Barbieri:2006dq}. 
In addition, those data can be parametrized through 
the electroweak oblique parameters $S$, $T$, and $U$~\cite{Peskin:1991sw}, 
and these three parameters are correlated to each other.  
Following by Ref.~\cite{Ilnicka:2015jba,Chun:2015hsa}, we can write down the form of $\chi^2_{\rm EWPT}$ as 
\begin{eqnarray}
\chi^2_{\rm EWPT}=
\begin{pmatrix} \Delta_S & \Delta_T & \Delta_U \end{pmatrix}
\begin{pmatrix}
    \sigma_{s}^2   &   C_{st}\sigma_{s}\sigma_{t}  & C_{su}\sigma_{s}\sigma_{u} \\
    C_{st}\sigma_{s}\sigma_{t}   &   \sigma_{t}^2  & C_{tu}\sigma_{t}\sigma_{u} \\
    C_{su}\sigma_{s}\sigma_{u}   &   C_{tu}\sigma_{t}\sigma_{u}  & \sigma_{u}^2 
\end{pmatrix}^{-1}
\begin{pmatrix} \Delta_S \\ \Delta_T \\ \Delta_U \end{pmatrix},
\end{eqnarray}
where the covariance matrix and bases $\Delta_S$, $\Delta_T$, and $\Delta_U$ 
are given by PDG data~\cite{Agashe:2014kda}. 
For covariance matrix elements, we use the values: 
$\sigma_{s}=0.1$, $\sigma_{t}=0.12$, $\sigma_{u}=0.1$, 
$C_{st}=0.89$, $C_{su}=-0.54$, and $C_{tu}=-0.83$. 
The bases are defined as $\Delta S= S-0.03$, $\Delta T=T-0.05$, and $\Delta U=U-0.03$.

\subsubsection{\bf Scalar bosons production at the LEP}
\label{sec:LEP}

Generally speaking, new scalar bosons ($S$, $A$ and $H^\pm$) 
can be produced either singly or doubly at the colliders.   
However, due to the protection of the extra $\mathbb{Z}_2$ symmetry in the i2HDM, 
all of these new scalar bosons can only be produced doubly. 
Therefore, those searches of single new scalar boson production in LEP, Tevatron and LHC 
cannot be applied in the i2HDM case. 
We review the searches for new scalar boson pair productions
at the LEP in the following.

First, if new scalar bosons are lighter than $W$ or $Z$ boson, the decay channels such as 
$W^\pm\to \{SH^\pm, AH^\pm\}$ and/or $Z\to \{SA,H^+H^-\}$ can be detected by LEP. 
Utilizing the null signal detections reported by LEP~\cite{Agashe:2014kda}, 
one can obtain 
\begin{eqnarray}
m_{S,A}+m_{H^\pm} &>& m_W, \nonumber\\
m_A+m_S &>& m_Z,\nonumber ~~{\rm and}~~\\ 
2 m_{H^\pm} &>& m_Z.
\end{eqnarray}
With the above criteria, the $W$ and $Z$ bosons cannot directly decay into these new scalar bosons.

Second, taking $S$ as the DM candidate, 
the CP-odd $A$ can decay into $S Z^{(\ast)}$, 
while the charged Higgs boson $H^\pm$ can decay into $W^{\pm(\ast)} S$.
If $H^\pm$ is heavier than $A$, the decay channel $H^\pm \to W^{\pm(\ast)} A \to  W^{\pm(\ast)} SZ^{(\ast)}$ can also be opened. 
Therefore, the final states of the two production processes $e^+e^-\to H^+H^-$ and 
$e^+e^-\to SA$ can be the signatures of missing energy together with multi-leptons or multi-jets, 
depending on the decay products of $W^\pm$ and $Z$ bosons. 
To certain extents, the signatures for charged Higgs searches in i2HDM 
can be similar to the supersymmetry searches for charginos at 
the $e^+e^-$ and hadron colliders~\cite{Aoki:2013lhm,Kalinowski:2018ylg,Dolle:2009ft}. 
Nevertheless, the cross sections for fermion and scalar boson pair productions 
are scaled by $\beta^{1/2}$ and $\beta^{3/2}$ respectively, 
where $\beta$ is the velocity of the final state particle in the center-of-mass frame. 
Hence, one can expect the production of the scalar pair is suppressed by an extra factor of $\beta$ 
compared with the fermionic case. 
The limits for a fermion pair (chargino-neutralino) production cannot directly applied on 
the scalar boson pair production such as $H^\pm H^\mp$ and $SH^\pm$~\cite{Pierce:2007ut}. 
In order to properly take the differences into account, 
we veto the parameter space based on the $95\%$ OPAL exclusion~\cite{Abbiendi:2003ji}. 
The exclusion has been recast and projected on ($m_{H^\pm}$, $m_A$) plane 
presented in Fig.~5 of Ref.~\cite{Blinov:2015qva}.

Finally, for $S A$ production mode, we can mimic the neutralino searches at LEP-II via 
$e^+e^- \to \widetilde{\chi}_1^0\widetilde{\chi}_2^0$ followed by 
$\widetilde{\chi}_2^0 \to \widetilde{\chi}_1^0 f\bar{f}$~\cite{Acciarri:1999km}. 
The process $e^+e^-\to SA$ followed by the cascade $A \to SZ^{(\ast)} \to S f\bar{f}$ can give similar signature 
and the detail analysis had been carefully done in Ref.~\cite{Lundstrom:2008ai}. 
In our approach, we use the exact exclusion region on ($m_S$, $m_A$) plane as given in Fig.~7 of 
Ref.~\cite{Lundstrom:2008ai} to veto the parameter space.

Since reconstructing these three LEP constraints with a precise likelihood 
will cost a lot of CPU-consuming computation, 
we only use hard-cuts to implement them into our analysis.   

\subsubsection{\bf Exotic Higgs decays}
\label{sec:inv}

Once these $\mathbb{Z}_2$ odd scalar bosons are lighter than a half of the SM-like Higgs boson $h$, 
the Higgs exotic decays $h\rightarrow S S/ A A/ H^+H^-$ can be opened. 
These exotic Higgs decays can modify the total decay width of the SM-like Higgs boson as well as 
the SM decay branching ratios which can be constrained by the current Higgs boson measurements  
and further tested by the future Higgs boson precision experiments. 
For the compressed mass spectrum scenario with the DM candidate $S$, 
the final states for $h\rightarrow S S$ are invisible while $h\rightarrow A A/ H^+H^-$ are 
missing energy plus very soft jets or leptons. 
In the case that these jets or leptons are too soft to be detected at the LHC, 
the signatures of $h\rightarrow A A/ H^+H^-$ 
are identical to $h\rightarrow S S$. 
Recently, both ATLAS and CMS have reported their updated limits on 
the branching ratio of Higgs invisible decays~\cite{Aaboud:2019rtt,Sirunyan:2018owy}.
Including the Higgs-strahlung $pp\to ZH/WH$ and 
the vector boson fusion (VBF) processes,
the ATLAS collaboration has reported an upper limit 
on the invisible branching ratio $\hinv<0.26$ at 95\% confidence level~\cite{Aaboud:2019rtt}.
Similarly, the CMS collaboration has also reported 
an upper limit on the invisible branching ratio $\hinv<0.19$ 
at 95\% confidence level~\cite{Sirunyan:2018owy} 
by the combining searches for Higgs-strahlung, VBF, 
and also gluon fusion (ggH) processes\footnote{
The CMS collaboration has reported their first search for the Higgs invisible decays 
via $t\overline{t}h$ production channel at $s=\sqrt{13}$ TeV~\cite{CMS:2019bke}, 
but the constraint $\hinv<0.46$ at 95\% confidence level
is much weaker than the combined one in Ref.~\cite{Sirunyan:2018owy}.}.
On the other hand, 
a recent global-fit analysis on the SM-like Higgs boson measurements using ATLAS and CMS data suggested 
a more aggressive constraint on the branching ratio for nonstandard decays of the Higgs boson 
to be less than 8.4\% at the 95\% confidence level~\cite{Cheung:2018ave}. 
In the near future, $\hinv$ is expected to reach the limit less than about 5\% at the HL-LHC~\cite{CMS:2018tip}.
For the sake of conservation, 
we only use the result from CMS ~\cite{Sirunyan:2018owy} in this analysis.


\subsubsection{\bf{Diphoton signal strength $R_{\gamma\gamma}$ in the i2HDM}}
\label{sec:diphoton}

Beside exotic Higgs decays, 
the rate of the SM-like Higgs boson decaying 
into diphoton can also be modified. 
In particular, the new contribution adding to the SM one is the charged Higgs triangle loop. 
Since, at the leading order, the couplings between the SM-like Higgs boson and SM particles are unchanged,
the production cross section of the Higgs boson will be the same as the SM one. 
Hence, we can obtain the diphoton signal strength 
in the i2HDM by normalized to the SM value: 
\begin{eqnarray}
R_{\gamma\gamma} \equiv  
\frac{\sigma_{h}^{\gamma\gamma}}{\sigma_{h_{\rm SM}}^{\gamma\gamma}} & \simeq &
\frac{{\rm BR}(h \to \gamma\gamma)^{\rm i2HDM} }
{{\rm BR}(h \to \gamma\gamma)^{\rm SM} }.
\label{ratio}
\end{eqnarray}
The exact formula for the partial decay width of $h\to\gamma\gamma$ 
in the i2HDM can be found in Ref.~\cite{Arhrib:2012ia,Swiezewska:2012eh} and 
${\rm BR}(h\rightarrow\gamma\gamma)^{\rm SM}=2.27\times 10^{-3}$ are taken from PDG data~\cite{Agashe:2014kda}. 
We apply the public code $\texttt{micrOMEGAs}$~\cite{Belanger:2018mqt} 
by using the effective operators as implemented in Ref.~\cite{Belyaev:2012qa} 
to calculate ${\rm BR}(h\rightarrow\gamma\gamma)^{\rm i2HDM}$ in this study. 
Recently, both ATLAS and CMS collaborations have reported their searches for 
the Higgs diphoton signal strength~\cite{ATLAS:2018doi,CMS:1900lgv}. 
In particular, a combined measurements of the Higgs boson production 
from ATLAS~\cite{ATLAS:2018doi} gives $R_{\gamma\gamma}= 1.08^{+0.13}_{-0.12}$. 
On the other hand, the measurements of Higgs boson production 
via ggH and VBF from CMS~\cite{CMS:1900lgv} give $R_{\gamma\gamma}=
1.15^{+0.15}_{-0.15}$ and $0.8^{+0.4}_{-0.3}$, respectively. 
One can see that all of these measurements are in agreement with the SM prediction. 
In this analysis, we only use the latest ATLAS result \cite{ATLAS:2018doi} 
to constrain the model parameter space.

\subsubsection{\bf{Mono-X and compressed mass spectra searches at the LHC}}

\underline{{\it Mono-jet:}}\\ 
One possible way to search for DM at the LHC 
is looking at final states with a large missing transverse energy 
associated with a visible particle such as 
jet~\cite{Aaboud:2017phn} and lepton~\cite{Aad:2019wvl}.
In the i2HDM, the mono-jet signal is a pair of DM  
produced by the Higgs boson and accompanied with at least one energetic jet.
If the pseudo-scalar $A$ has a small mass splitting with the DM 
and decays into very soft and undetectable particles, 
it can also contribute to the mono-jet signature. 
For the case of $m_S > m_h/2$, since the DM pairs are produced through an off-shell Higgs, 
the cross section of the mono-jet process is suppressed. 
On the other hand, 
for $m_S < m_h/2$ case, the missing transverse energy is usually low 
that the mono-jet constraint is less efficient.

We recast the current ATLAS mono-jet search~\cite{Aaboud:2017phn} 
by using \texttt{Madgraph 5}~\cite{Alwall:2014hca} and \texttt{Madanalysis 5}~\cite{Dumont:2014tja}. 
It turns out that the current search excludes $\lambda_S > 3\times 10^{-2}$ 
for the case of $m_S < m_h/2$, and $\lambda_S > 5.0$ for higher DM masses. 
We will see later that these limits are much weaker than other DM constraints.

\underline{\it Mono-lepton:} \\
The mono-lepton signal in this model is raised from the process
$p p \to S H^{\pm}$ with $H^{\pm} \to S l^+ \nu$. 
However, the current mono-lepton search from ATLAS~\cite{Aad:2019wvl} 
is not really sensitive to the small mass splitting $\Delta^\pm$. 
Indeed, the signal efficiency is too low to be detected because the transverse mass distribution of 
the lepton and missing transverse momenta in the final state is not large enough. 
Therefore, this constraint cannot be applied in this work. 

\underline{\it Compressed mass spectra search:}\\
Searches for events with missing transverse energy and two same-flavor, opposite-charge, 
low transverse momentum leptons 
have been carried out 
from CMS~\cite{Sirunyan:2019zfq}
and ATLAS~\cite{Aaboud:2017leg,Aad:2019qnd} Collaborations. 
These typical signatures are sensitive to any model 
with compressed mass spectra if the production cross section is large enough.
In the i2HDM, 
the pairs of $AS$, $AH^{\pm}$ and $H^{\mp}H^{\pm}$ 
can be produced at the LHC via the 
$q\bar{q}$ fusion and VBF processes.
The heavier scalar $A$ then can 
decay into a dilepton pair 
via an off-shell $Z$ boson, 
such that the dilepton invariant mass ($m_{ll}$) 
is sensitive to the mass-splitting $\Delta^0$. 
On the other hand, the charged Higgs $H^{\pm}$ 
can decay into a lepton and a neutrino
via an off-shell $W$ boson. 
The stransverse mass $m_{T2}$
is sensitive to the mass-splitting $\Delta^{\pm}$.

We recast the ATLAS SUSY compressed mass spectra search~\cite{Aad:2019qnd}. 
The matrix element generator \texttt{Madgraph~5}~\cite{Alwall:2014hca} is used 
to generate the signal events at leading order
which are then interfaced with \texttt{Pythia~8} \cite{Sjostrand:2014zea} 
for showering and hadronization, 
and \texttt{Delphes~3} \cite{deFavereau:2013fsa} 
for the detector simulations. 
The \texttt{Madanalysis~5} package~\cite{Dumont:2014tja}
is used to recast the experiment results.
We apply the same preselection requirements 
and signal regions selection cuts as in Ref.~\cite{Aad:2019qnd}.
Two opposite-charged muons in the final states are chosen for recasting in this study.
In our parameter space of interest, 
a suitable signal region is the one labeled as \textbf{SR-E-low} in Ref.~\cite{Aad:2019qnd} 
with the muon pair invariant mass window: 3.2 GeV $\leq m_{\mu^+\mu^-} \leq$ 5 GeV. 
Due to the small 
production cross section of 
$AS$, $AH^{\pm}$ and $H^{\mp}H^{\pm}$,
the current data at the LHC cannot probe 
the parameter space of interest,  
but the sensitivity at future HL-LHC may be expected to 
reach it. 

\subsection{Relic density}
\label{sec:relic}

\begin{figure}[htbp]
\begin{centering}
\includegraphics[width=0.6\textwidth]{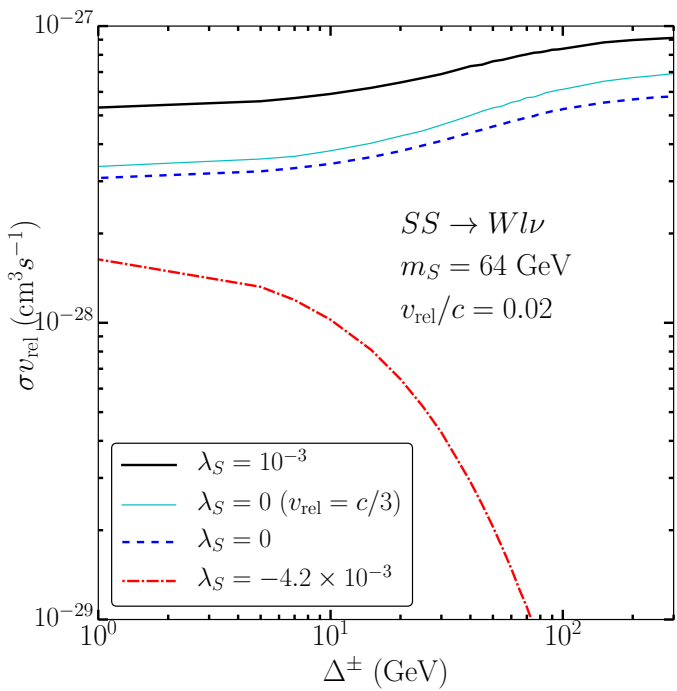}
\caption{
The cross section $\sigma v_{{\rm rel}}$ of $SS\to WW^\ast\to W l\nu$ as a function of $\Delta^\pm$. 
The mass of DM is fixed to be $m_S=64\gev$. 
Except the cyan line ($v_{{\rm rel}}=c/3$), we take a relative velocity $v_{{\rm rel}}=0.02 c$ 
which can make a cancellation effect for $\lambda_S = -(s-m_h^2)/(2v^2) \simeq -4.2 \times 10^{-3}$ 
as discussed in the main text. 
The black solid line, blue dashed and red dashed-dotted lines 
represent the values of $\lambda_S = 10^{-3}$, $\lambda_S= 0 $ and $\lambda_S = -4.2\times 10^{-3}$, respectively.
}
\label{fig:xxww}
\end{centering}
\end{figure}

Assuming a standard thermal history of our Universe, the number density of a particle with mass $m$
at the temperature $T$ can be simply presented by a Boltzmann distribution $\sim e^{-m/T}$. 
Under the thermal DM scenario, if the mass splittings between $S$, $A$, 
and $H^\pm$ are as small as the case 
considering in this analysis, the number densities of three particles at $T$ are comparable 
to each other and their co-annihilating processes play an important role 
of reducing the relic density. 
In this subsection, we summarize the dominant channels for annihilation 
and co-annihilation in the i2HDM.

Depending on the specific DM mass range, the DM annihilation is dominated by different channels. 
For the Higgs resonance regions, \textit{i.e.} $m_S\simeq m_h/2$, 
the annihilations of $SS$ to SM fermions via the Higgs boson exchange are dominated, 
especially for the process $SS\rightarrow b\overline{b}$. 
The annihilation cross section in the function of the central energy $\sqrt{s}$
is given by 
\begin{eqnarray}
\label{eq:hres}
\sigma(s)&=&\frac{3\lambda_S^2}{8\pi} \frac{m_f^2}{\left(s-m_h^2\right)^2+\Gamma_h^2 m_h^2}
\times r^3(s,2 m_f) \times r(s,2 m_S),~~\rm{where}\nonumber \\
r(x,y) &\equiv & \sqrt{1-\frac{y^2}{x^2}}.  
\end{eqnarray}
It is easy to see that the annihilation cross section from Eq.~(\ref{eq:hres}) 
is dramatically decreased when $m_S>m_h/2$.

For the region of $m_S>m_h/2$, 
the annihilation of $S S$ to $W^+ W^{-(\ast)}$
via the four-vertex interaction, the $s$-channel $h$ exchange, and the $t$ and $u$-channel with $H^\pm$ exchange,
becomes the dominant channel.
The contribution from four-vertex diagram is usually dominant 
while the one from charged Higgs is typically small. 
The one from $h$ exchange becomes important only at around the Higgs resonance region.  
However, the total $S S\to W^+ W^{-(\ast)}$ annihilation can be significantly 
reduced by the cancellations among these contributions. 
First, the cancellation between the four-vertex diagram and $h$ 
exchange contribution can take place at $\lambda_S = -(s-m_h^2)/(2v^2)$ \cite{Honorez:2010re}. 
Second, if the masses of $S$ and $H^\pm$ are nearly degenerate, 
the cancellation between the four-vertex diagram and charged Higgs contribution 
can also occur.
In Fig.~\ref{fig:xxww}, we show the cross section times relative velocity ($\sigma v_{{\rm rel}}$) 
for the process $SS\to WW^\ast\to W l\nu$ 
as a function of $\Delta^{\pm}$ with various values of $\lambda_S$. 
The cross section is computed by using \texttt{MadGraph5}. 
The DM mass is fixed to be $m_S=64\gev$ and 
its relative velocity in the early universe is taken as $v_{{\rm rel}}/c=0.02$ 
in order to realize a cancellation between the four-vertex diagram and $h$ exchange contribution 
for $\lambda_S = -(s-m_h^2)/(2v^2) \simeq -4.2 \times 10^{-3}$. 
We also present the cyan line ($\lambda=0$ with $v_{{\rm rel}}=c/3$) as a reference.  
Three benchmark values of $\lambda$ are selected based on the allowed region given 
in a previous study~\cite{Arhrib:2013ela}. 
For a positive value of $\lambda_S = 10^{-3}$ (black solid line), 
the contribution from the four-vertex diagram and s-channel Higgs exchange are dominant. 
When the $s$-channel Higgs exchange is removed by setting $\lambda_S=0$ (blue dashed line), 
$\sigma v_{{\rm rel}}$ is reduced but still sizable. 
For $\lambda_S = -4.2\times 10^{-3}$ (red dashed-dotted line), 
one can see that $\sigma v_{{\rm rel}}$ is the smallest 
and even can be neglected if the charged Higgs mass becomes heavy. 
This is due to the maximal cancellation between the four-vertex diagram and $h$ exchange contribution. 
The remaining $\sigma v_{{\rm rel}}$ is mainly from the charged Higgs contribution.
One can also see that from the black solid and blue dashed lines, 
$\sigma v_{{\rm rel}}$ is slightly dropped if the mass splitting $\Delta^\pm$ decreases from 
$\sim 250$ GeV down to $\sim 1$ GeV. 
This is the result of the cancellation between the four-vertex diagram and $t/u$ channel charged Higgs contributions.

Unlike the most parameter space of the i2HDM, the difficulty of a compressed mass spectrum scenario 
is that additional co-annihilation channels in this scenario make under-abundant relic density.
Ideally, the annihilation has to be properly switched off, but it is particularly hard  
for $SS\to WW^{(\ast)}$ because the coupling of four-vertex diagram $SSWW$ is the electroweak coupling.  
Therefore, the only way to reduce the annihilation cross section of $SS\to WW^*$ 
is via cancellation as shown in Fig.~\ref{fig:xxww}. 
Since the mass splitting $\Delta^\pm$ in this work is always greater than $1\gev$, 
it is interesting to see the role of the $t/u$-channel with $H^\pm$ exchange playing in the cancellation. 
By engaging with \texttt{micrOMEGAs} package\footnote{
All the off-shell contributions have been implemented in the \texttt{micrOMEGAs} in this work.},  
we have checked that if we change $\Delta^\pm$ from $20\gev$ to $500\gev$, 
the relic density is slightly changed by a few percent. 
Hence, the $t/u$-channel with $H^\pm$ exchange are not the leading contribution 
to suppress the contribution from $SS\to WW^\ast$ annihilation in the co-annihilation scenario.

The co-annihilation contribution to the relic density is more complicated than the annihilation process. 
In particular, it depends on the size of mass splitting $\Delta^0$ and $\Delta^\pm$. 
Since we are focusing on the compressed mass spectrum scenario, 
the co-annihilation happens naturally and cannot be expelled from the full mass regions. 
In such a small mass splitting scenario, the most dominant co-annihilation channels are:   
\begin{itemize}
    \item $SA\to f\bar{f}$ for a small $\Delta^0$.      
    \item $S H^\pm \to \gamma W^\pm$ for a small $\Delta^\pm$. 
\end{itemize}
Unlike other subdominant co-annihilation channels, 
these two co-annihilation cross sections 
are only involved with the SM couplings.
Naively speaking, the relic density in the co-annihilation dominant region is essentially 
controlled by the two mass splittings $\Delta^0$ and $\Delta^\pm$.

We evolve the Boltzmann equation by using the public code \texttt{MicrOMEGAs}~\cite{Belanger:2018mqt}. 
The numerical result of the relic density 
which has been taken into 
account the annihilation and co-annihilation contributions, 
is required to be in agreement 
with the recent PLANCK measurement~\cite{Aghanim:2018eyx}:
\begin{equation}
\Omega_{\rm{CDM}} h^{2}=0.1199\pm0.0027.
\label{Omega}
\end{equation}

We would like to comment on the multi-component DM within the framework of the i2HDM. 
If there exists more than one DM particle in the Universe, the DM $S$ can be only a fraction of 
the relic density and the DM local density. The DM constraints from relic density, direct detection, and indirect detection 
can be somewhat released. 
However, an important question followed by adding more new particles to the Lagrangian 
is whether the Higgs potential is altered and theoretical constraints are still validated. 
Such a next-to-minimal i2HDM is indeed interesting but beyond the scope of our current study. 
Here, we only consider the one component DM scenario.

\subsection{\bf DM direct detection}
\label{sec:DD}
As indicated by several Higgs portal DM models~\cite{Cheung:2012xb,Athron:2017kgt,Athron:2018hpc}, 
the most stringent constraint on DM-SM interaction currently comes from the DM direct detection. 
This is also true in the case of the i2HDM, if we only consider the DM-quark/gluon 
elastic scattering via $t$-channel Higgs exchange at the leading order. 
Simply speaking, one would expect that 
the size of $S-S-h$ effective coupling ($\lambda_{S}$ at tree-level)
can be directly constrained by the latest XENON1T experiment~\cite{Aprile:2018dbl}.

However, for a highly mass degenerated scenario $\Delta^0\simeq 0$, 
the DM-quark inelastic scattering $S q\to A q$ can be 
described by an unsuppressed coupling $Z-S-A$ 
whose size is fixed by the electroweak gauge coupling. 
Indeed, such the DM inelastic scattering scenario 
predicts a huge cross section but it has already excluded 
by the latest XENON1T result~\cite{Arina:2009um,Chen:2019pnt}.
The inelastic interactions are inefficient when 
the $\Delta^0$ is larger than the momentum exchange $\simeq \mathcal{O}(200\kev)$~\cite{Arina:2009um}.
Hence, it is safe to ignore the inelastic interactions 
when requiring $\Delta^0\geq 10^{-3}\gev$.
We not that $\Delta^0$ has to be greater than $5\gev$ 
when the DM relic density constraint is considered.

Going beyond the leading order calculation, 
we calculate the corrections at the next leading order in the i2HDM. 
We fold all the next leading order corrections into the effective coupling $\lambda_S^{\rm eff}$
which depends on the relative energy scale we have set.
Once the input scales of our scan parameters are fixed at the EW scale, 
the effective coupling $\lambda_S^{\rm eff}$ will be modified at the low energy scale 
where the recoil energy of DM-quark/gluon scattering is located. 
As shown in the Appendix~\ref{app:sigsip_NLO}, 
the one-loop correction $\delta\lambda$ is a function of $m_S, \Delta^0, \Delta^{\pm}$, 
and $\lambda_2$. 
Its value can be either positive or negative. 
Therefore, we can introduce a factor $R$ to illustrate the loop-induced effects, 
\begin{equation}
R=\frac{\sigsip(\texttt{tree+loop})}{\sigsip(\texttt{tree})}= 
\left(\frac{\lambda_{S}+\delta\lambda}{\lambda_{S}}\right)^2=
\left(\frac{\lambda_S^{\rm eff}}{\lambda_{S}}\right)^2.
\label{eq:NLORatio}
\end{equation} 
The correction parameter $\delta\lambda$ is computed by using \texttt{LoopTools} code~\cite{Hahn:1998yk}.
We consider two scenarios in this work:   
the XENON1T likelihood (Poisson distribution) is obtained with and without $R$ 
by using \texttt{DDCalc} code~\cite{Workgroup:2017lvb}.
Note that both the tree and loop level are isospin conserving. 
Hence, the value of $R$ for DM-proton and DM-neutron scattering are approximately the same.

\subsection{\bf DM indirect detection}
\label{sec:ID}
In addition to the singlet Higgs DM whose annihilation is only via SM Higgs exchange, 
the i2HDM at the present universe can be also dominated by the four-points interaction $SSW^+W^-$.
If the DM annihilation is considerable, 
e.g. at dwarf spheroidal galaxies (dSphs) or galactic center where DM density is 
expected to be rich, some additional photons or antimatter produced by $W$ bosons or SM fermion pair 
would be detected by the DM indirect detection. 
Unfortunately, none of the indirect detection experiments 
reports a positive DM annihilation signal but gives a sever limit on the DM annihilation cross section. 
Thanks to a better measured dSphs kinematics which gives smaller systematic uncertainties than other DM indirect detection,  
the most reliable limit on the DM annihilation cross section at the DM mass $\sim \mathcal{O}(100\gev)$ 
comes from Fermi dSphs gamma ray measurements so far~\cite{Fermi-LAT:2016uux}.

On the other hand, several groups have found out 
some anomalies such as GCE~\cite{Goodenough:2009gk,Hooper:2010mq,Calore:2014xka,TheFermi-LAT:2017vmf} 
and AMS02 antiproton excess~\cite{Cui:2016ppb,Cuoco:2016eej} 
which might be able to be explained by the DM annihilation. 
Interestingly, these two anomalies are located at the DM mass $\sim 50-100\gev$ region 
where coincides with the mass region discussed in this paper.    
Hence, we adopt a strategy in this work that only Fermi dSphs gamma-ray constraints are included 
in the scan level, but our allowed parameter space is compared with 
antiproton anomaly.

The differential gamma-ray flux due to the DM annihilation at the dSphs halo 
is given by  
\begin{equation}
\frac{d\Phi_\gamma}{dE_{\gamma}}={\frac{\sigmav}{8{\pi}m_{S}^{2}} \times J\times 
\sum_{\texttt{ch}} {\rm BR}(\texttt{ch})\times \frac{dN^{\texttt{ch}}_{\gamma}}
{dE_{\gamma}}}.
\end{equation}
The $J$-factor is $J={\int}dl d{\Omega}{\rho}(l)^{2}$, where 
the integral is taken along the line of sight from the detector 
with the open angle $\Omega$ and the DM density distribution $\rho$. 
We adopt 15 dSphs and their $J$-factors as implemented in \texttt{LikeDM}~\cite{Huang:2016pxg}. 
We sum over all the DM annihilation channels $\texttt{ch}$. 
The annihilation branching ratio ${\rm BR}(\texttt{ch})$ and energy spectra
$dN^{\texttt{ch}}_{\gamma}/dE_{\gamma}$ are computed by using
\texttt{micrOMEGAs} 
in which the three-body final states (e.g. $SS\to WW^*\to W^+ l^-\nu$) 
are properly taken into account.
In this paper, we only focus on the region of $m_S<100\gev$. 
For the DM indirect detection at the region of $m_S>100\gev$, 
the future Cherenkov Telescope Array may give a severe limit~\cite{Queiroz:2015utg}.


\section{Results and discussions \label{section:result}}



\subsection{Numerical method~\label{section:method}}


\begin{table}[t!]
    \centering
    \begin{tabular}{l|l|l}
        Likelihood type & Constraints & See text in\\
        \hline
        Step &  perturbativity, stability, tree-level unitarity & Sec.~\ref{section:theo}\\
        & LEP-II, OPAL & Sec.~\ref{sec:LEP}\\ 
        Poisson & XENON1T (2018), Fermi dSphs $\gamma$ data  & Sec.\ref{sec:DD}, \ref{sec:ID}\\
        Half-Gaussian &  exotic Higgs decays & Sec.~\ref{sec:inv}\\
        Gaussian & relic abundance, $R_{\gamma\gamma}$, EWPT  
        & Sec.~\ref{sec:relic}, \ref{sec:relic}, \ref{sec:EWPT}\\
        \hline
    \end{tabular}
    \caption{Likelihood distributions used in our analysis.}
    \label{tab:likelihood}
\end{table}

In a similar procedure developed in our previous works~\cite{Arhrib:2013ela,Cheung:2014hya,
Matsumoto:2014rxa,Matsumoto:2016hbs,Banerjee:2016hsk,Matsumoto:2018acr}, 
we use the likelihood distribution given in the Table~\ref{tab:likelihood} 
in our Markov Chain Monte Carlo scan.
Considering the lower DM mass 
which might be potentially detected in 
the colliders, DM direct and indirect detections,  
we only focus on the DM mass less than $100\gev$.
Engaging with \texttt{emcee}~\cite{ForemanMackey:2012ig}, 
we perform 35 Markov chains in the five dimensional parameter space, 
\begin{eqnarray}
\label{domain}
5.0 &\leq \, m_S  / \textrm{ GeV} \, \leq & 100.0 \; , \nonumber\\
10^{-3} &\leq \, \Delta^0  / \textrm{ GeV} \, \leq & 20 \; , \nonumber\\
1.0 &\leq \, \Delta^{\pm}  / \textrm{ GeV} \, \leq & 30 \; ,\nonumber\\
-2.0 &\leq \, \lambda_S  \, \,\leq& 2.0 \; , \nonumber\\
0.0 &\leq \, \lambda_2  \,\leq& 4.2 \; . \nonumber
\end{eqnarray}
Here, we only choose $\Delta^\pm$ up to $30\gev$ but one can freely extend it to a larger value until 
$m_{H^{\pm}}\sim 250\gev$ disfavored by the EWPT data  
(for a detailed analysis, see Ref.~\cite{Bhardwaj:2019mts}).
On the other hand, without the helps from the charged Higgs co-annihilations, 
we have checked that the total co-annihilation contribution to the relic density 
cannot be larger than $85\%$.

In order to scan the parameter space more efficiently, we set the range of 
$\lambda_2$ up to 4.2 allowed by the unitarity constraint~\cite{Arhrib:2013ela}. 
We compute the mass spectrum, theoretical conditions and oblique parameters 
by using \texttt{2HDMC}.  
All survived parameter space points are then passed to 
\texttt{micrOMEGAs} 
to compute the relic density and the tree-level DM-nucleon elastic scattering cross section. 
The XENON1T statistics test is computed by using \texttt{DDCalc}. 
The loop corrections of DM-nucleon elastic scattering cross section 
is calculated by using \texttt{LoopTools}.    
In the end, we have collected more than 2.5 million data points and  
our achieved coverage of the parameter space is good enough to 
pin down the contours by using ``\textit{Profile Likelihood}'' method~\cite{Rolke:2004mj}. 
Under the assumption that all uncertainties 
follow the approximate Gaussian distributions, 
confidence intervals are calculated from the tabulated values of 
$\Delta\chi^2\equiv -2\ln(\mathcal{L/L}_{\rm{max}})$. Thus,
for a two dimension plot, the 95\% confidence ($2\sigma$)
region is defined by $\Delta\chi^2 \leq 5.99$.

\subsection{Co-annihilation and mixed scenarios}

\begin{figure}[htbp]
\begin{centering}
\includegraphics[width=0.45\textwidth]{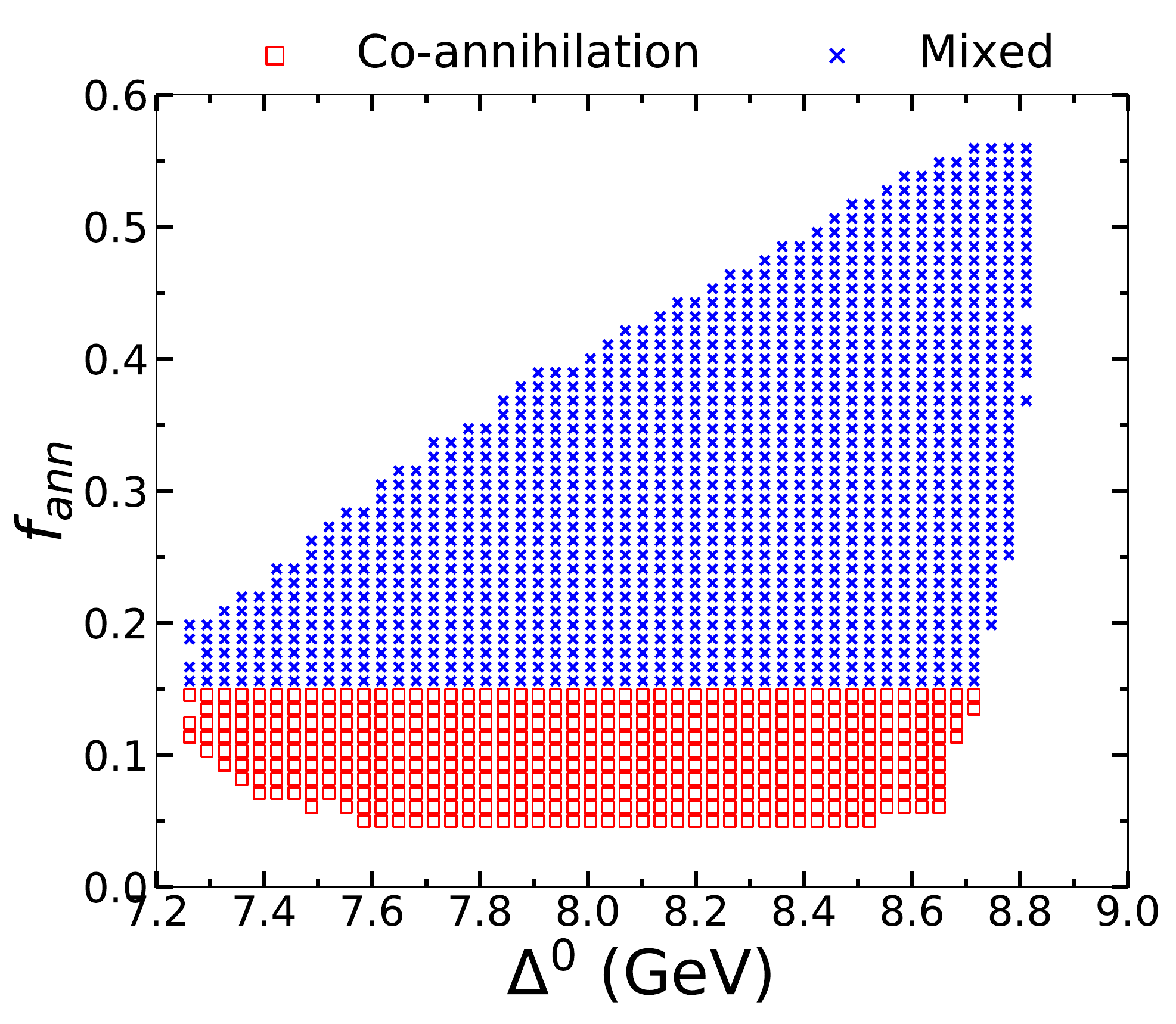} 
\includegraphics[width=0.45\textwidth]{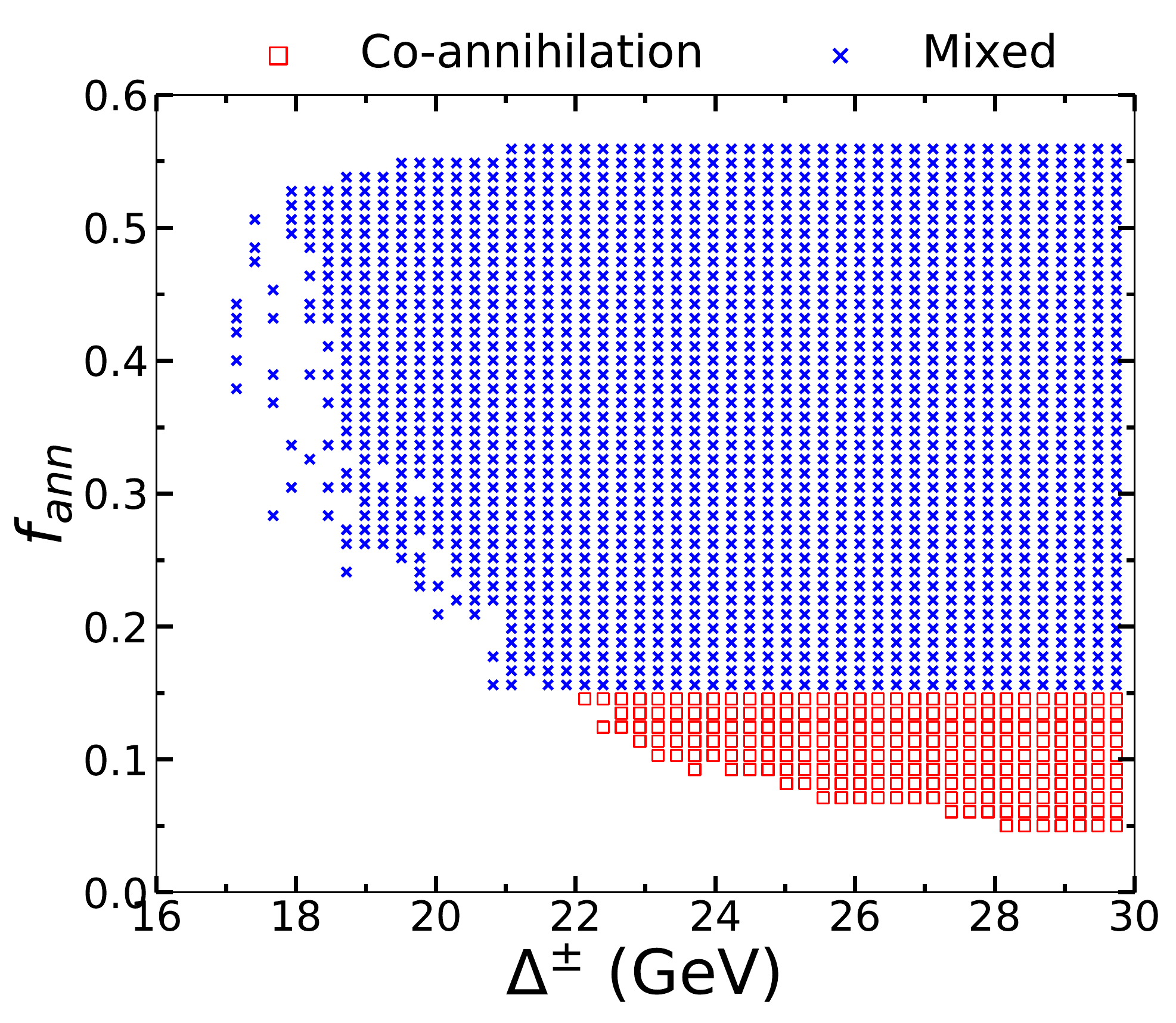}\\
\includegraphics[width=0.45\textwidth]{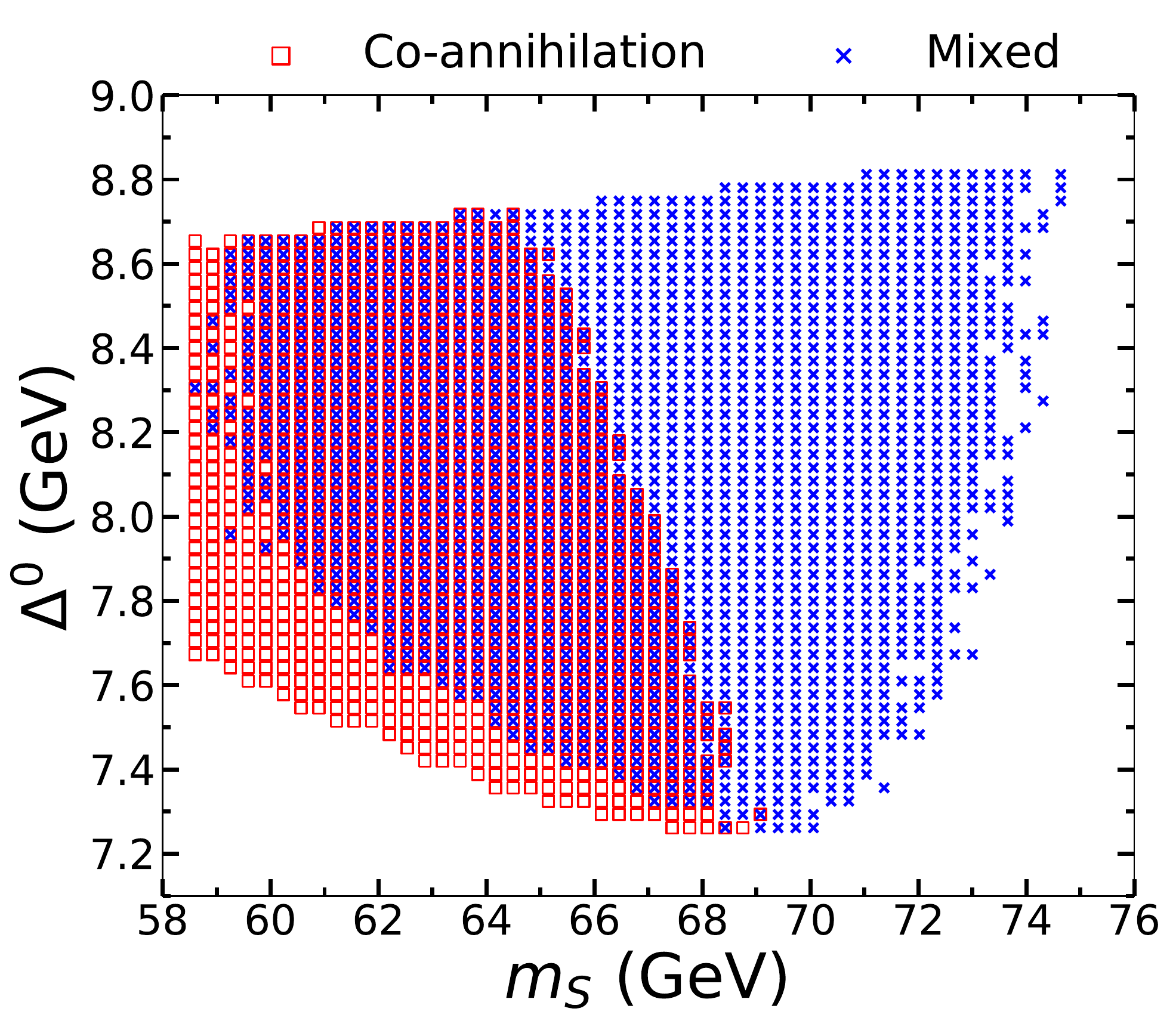}
\includegraphics[width=0.45\textwidth]{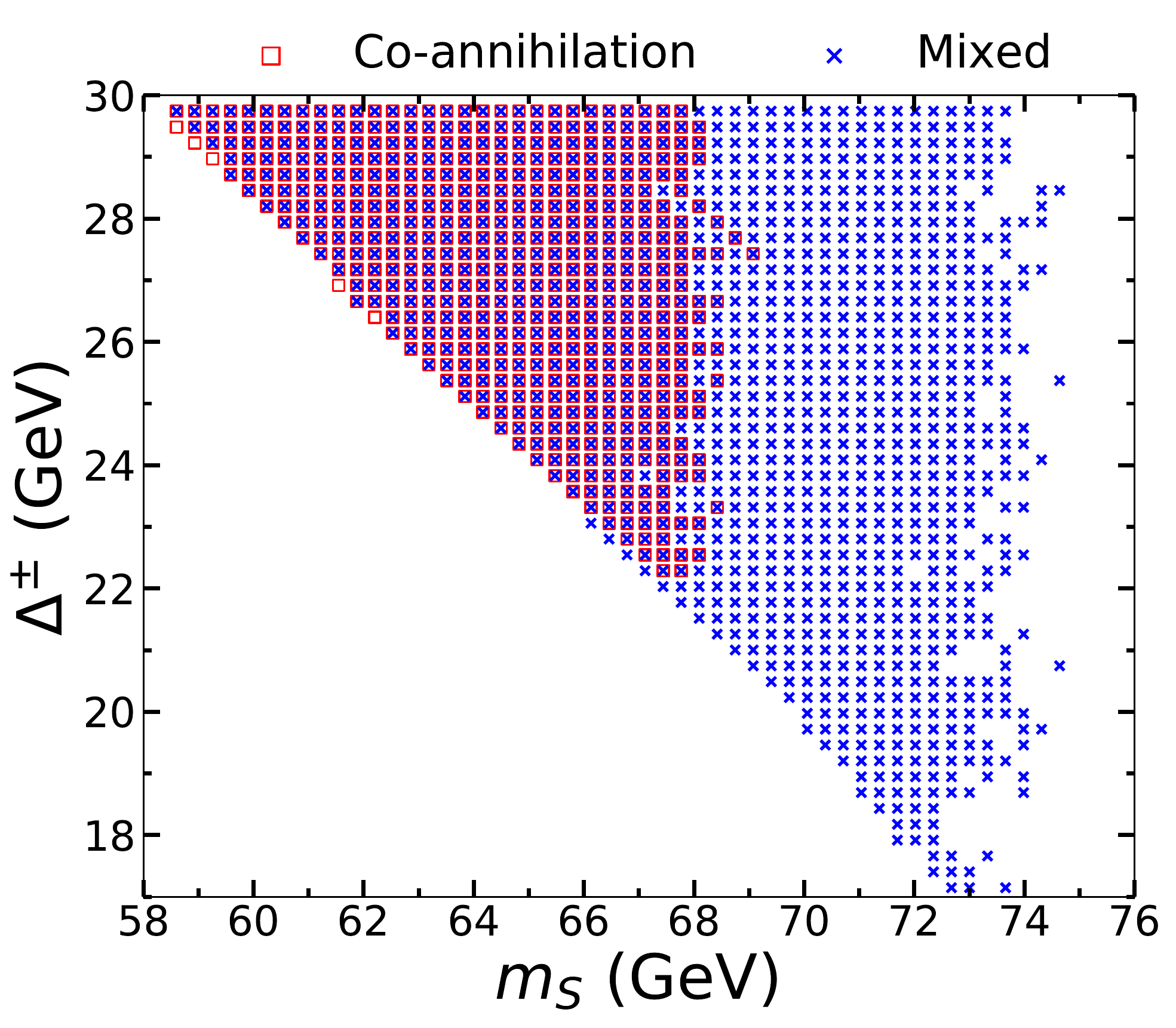}\\
\caption{
The $2\sigma$ allowed regions from 
the theoretical conditions, LEP, 
and PLANCK constraints.
The red square and blue cross 
are the co-annihilation ($f_{ann} < 15\%$) 
and mixed regions ($f_{ann} > 15\%$), respectively}
\label{fig:relic-lowmass}
\end{centering}
\end{figure}

As mentioned in Sec.~\ref{sec:relic}, 
the compressed mass spectrum scenario 
guarantees the number density of $S$, $A$ and $H^{\pm}$  
at the same temperature before freeze-out are comparable.  
Nevertheless, the thermal averaged cross section of co-annihilation 
could be very different with annihilation one.
Therefore, the interaction rate can describe the effects 
from both the thermal averaged cross section and their mass splitting.
If the interaction rate is below the Hubble expansion rate, 
the freeze-out mechanism occurs. 
Additionally, it is hard to cease the annihilation processes, 
particularly from four-points interactions
whose the contributions are always significant 
at the mass region $m_h/2<m_S\lesssim 500\gev$.
Here, we introduce a new parameter $f_{ann.}$ 
to account for the fraction of interaction rate  
attributable to the annihilation. 
Following the convention from \texttt{micrOMEGAs}, 
the fraction $f_{ann.}$ can be given by
\begin{eqnarray}
\label{eq:fann}
f_{ann.}&=&\frac{\Gamma_{ann.}}{\Gamma_{tot.}},
\end{eqnarray}
where $\Gamma_{ann.}$  and $\Gamma_{tot.}$ 
are the annihilation and total interaction rate 
before freeze-out, respectively.
In this analysis, we assume that 
the co-annihilation domination before freeze-out 
acquires $f_{ann.}< 0.15$.

The allowed regions of theoretical conditions, 
LEP, and PLANCK $2\sigma$ 
projected on ($\Delta^0$, $f_{ann.}$) and ($\Delta^\pm$, $f_{ann.}$) planes 
are shown in the two upper panels of Fig.~\ref{fig:relic-lowmass}.
We found that the co-annihilation rate ($\Gamma_{tot.}-\Gamma_{ann.}$) 
is always larger than $45\%$ in our parameter space of interest.
Therefore, there is no pure annihilation contribution
in this compressed mass spectrum scenario.
We also see that the relic density constraints 
can give a lower limit 
for both $\Delta^0$ and $\Delta^\pm$ 
because the co-annihilation processes are too 
efficient to reduce the relic density. 
On the other hand, $\Delta^0 < 8.7\gev$ 
is caused by the LEP-II constraints. 
We note that there is no upper limit of $\Delta^\pm$ 
from neither OPAL exclusion nor PLANCK measurement. 
The small $m_S$ and $\Delta^\pm$ regions are excluded by the OPAL 
as shown in the right bottom panel of Fig.~\ref{fig:relic-lowmass}.

Once the phase space of $SS\to W^+W^{-(\ast)}$ process is suppressed at the early universe, 
the co-annihilation rate dominates $\Gamma_{tot}$ before freeze-out. 
As a result, the co-annihilation scenario $f_{ann.} < 0.15$ only located at the region $m_S<70\gev$.  
The annihilation process $SS\to W^+W^{-(\ast)}$ is still at least $5\%$ 
contribution to $\Gamma_{tot}$.

There are three edges of the allowed region in ($m_S$, $\Delta^0$) plane. 
The left-bottom and right-bottom corner are excluded by 
too little and too much relic density, respectively.
The upper limit on the mass splitting, 
$\Delta^0 \lsim 8.8\gev$, is due to the LEP limit and 
this also leads to that $f_{ann.}\lesssim 55\%$.
Therefore, the expected features of annihilation are hidden in the region of the Mixed scenario. 
For example, near the Higgs resonance region ($m_S\simeq 62\gev$), 
one can see a small kink at the edge $\Delta^0\approx 7.6-7.7\gev$ of the Mixed scenario region.

\subsection{Results}

\begin{figure}[htbp]
\begin{centering}
\includegraphics[width=0.45\textwidth]{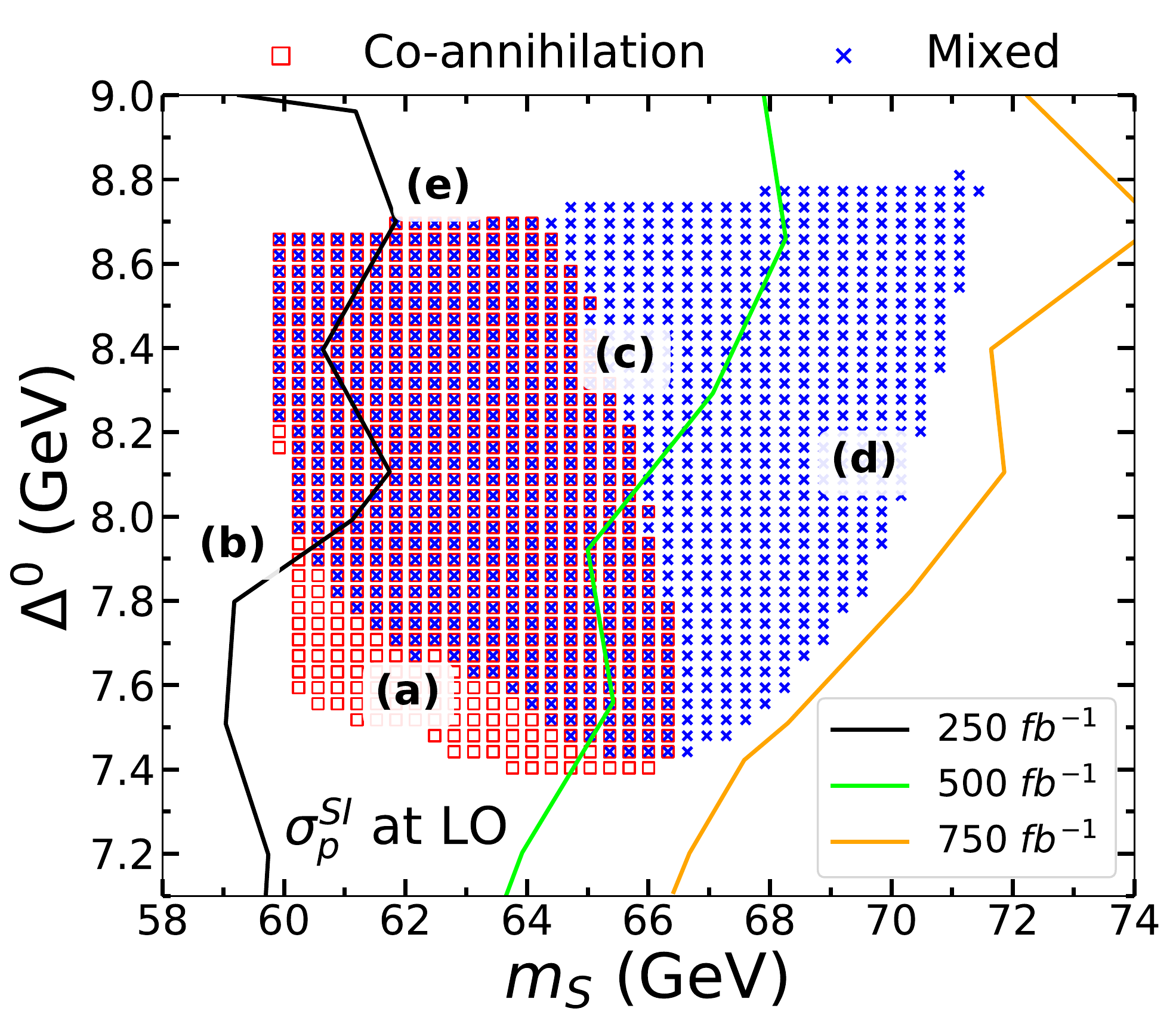}
\includegraphics[width=0.45\textwidth]{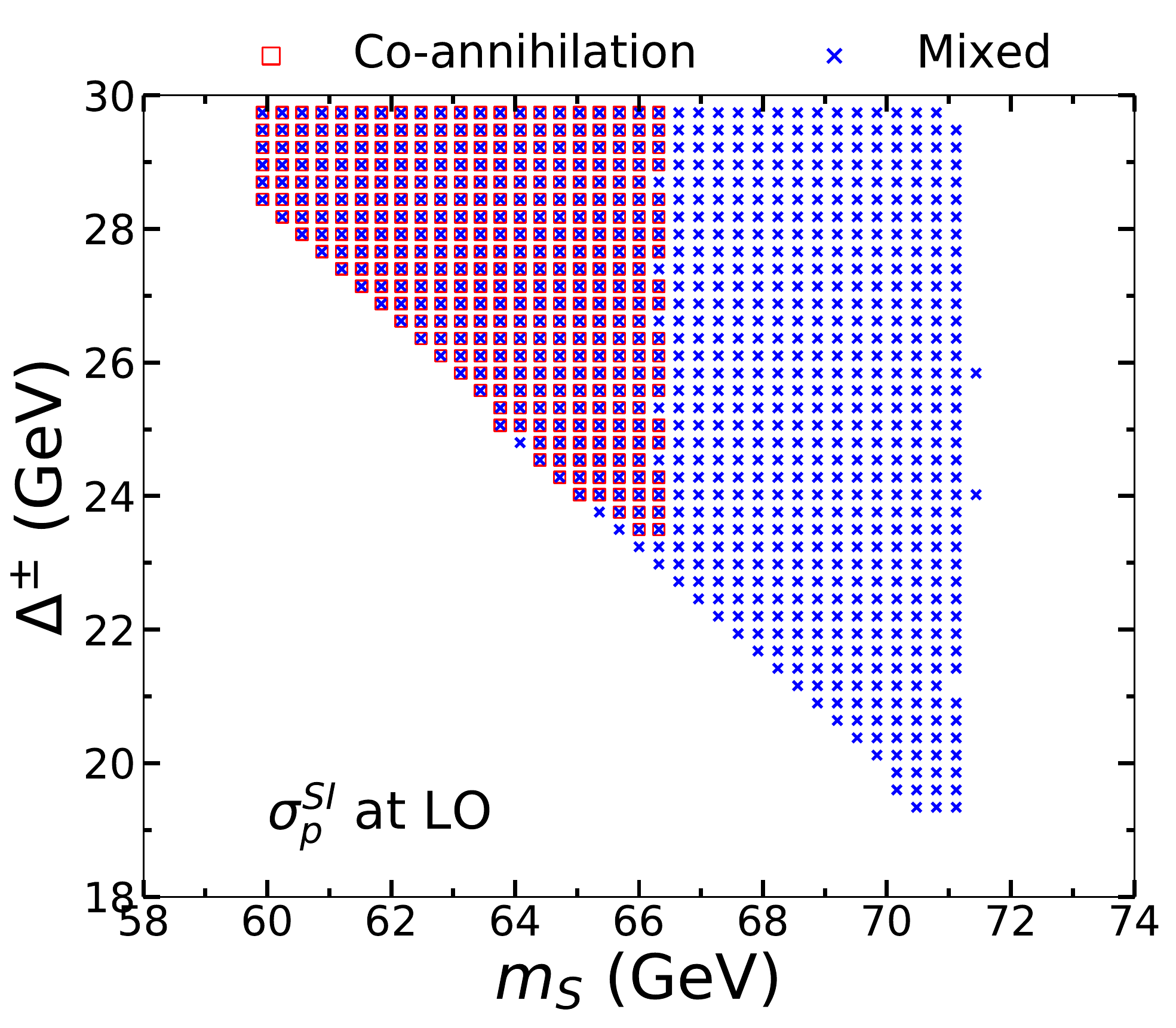}
\caption{
Left panel: the $2 \sigma$ allowed regions 
from all constraints
projected on ($m_S$, $\Delta^0$) plane. 
The black, green, and orange contours 
are the prospect of $2\sigma$ significance with the integrated luminosity of 
250 ${\rm fb}^{-1}$, 500 ${\rm fb}^{-1}$, and 750 ${\rm fb}^{-1}$ 
for one of the signal regions in Ref.~\cite{Aad:2019qnd} as shown in the main text, respectively. 
Right panel: the $2 \sigma$ allowed regions 
from all constraints
projected on ($m_S$, $\Delta^{\pm}$) plane. 
The $\sigsip$ is calculated at leading order and the colored codes for the scatter points are the same as Fig.~\ref{fig:relic-lowmass}. 
}
\label{fig:fig2}
\end{centering}
\end{figure}

In this section, we present the $2\sigma$ allowed region based on the total likelihoods, 
as shown in Table~\ref{tab:likelihood}. 
Those constraints have been discussed
in Sec.~\ref{sec:constraints} and they are referred to   
the phrase ``\textbf{all constraints}", unless indicated otherwise.

Fig.~\ref{fig:fig2} shows the $2 \sigma$ allowed region 
by taking into account all constraints. 
The left panel is on the ($m_S$, $\Delta^0$) plane 
and the right panel is on ($m_S$, $\Delta^{\pm}$) plane. 
Here, the $\sigsip$ is computed at LO. 
By comparing with the lower panels in Fig.~\ref{fig:relic-lowmass},
we can easily see that 
the allowed range of the DM mass is shrunk to $60 < m_S/\gev < 72$ 
after applying all constraints. 
Since the correlations between $m_S$, $\Delta^0$, and $\Delta^\pm$ 
are non-trivial 
but important for 
understanding the different features of mixed scenario (blue crosses) 
and co-annihilation scenario (red squares), 
we label some specific regions shown in the left panel of Fig.~\ref{fig:fig2} and discuss some features for these regions as follows:
\begin{itemize}
\item[]\textbf{Region (a)}, 
the small $\Delta^0$ can provide an efficient co-annihilation 
to reach the correct relic density even without the annihilation. 
One has to keep in mind that 
the coupling $\lambda_S$ can be either very small or a negative value in this region. 
It depends on whether $SS\to W W^{\ast}$ is close (a tiny $\lambda_S$) or opened (a negative $\lambda_S$).

\item[]\textbf{Region (b)},
comparing with the lower panel in Fig.~\ref{fig:relic-lowmass}, 
involving the Higgs invisible decay 
constraint lifts the lower limit on DM mass to $60\gev$.

\item[]\textbf{Region (c)},
the co-annihilation scenario cannot reach the large $m_S$ regions, 
particularly $m_S < 64$ GeV for $\Delta^0 = 8.7$ GeV 
and $m_S < 66\gev$ for $\Delta^0 = 7.4\gev$.
This is due to the current DM direct detection constraint.
In particular, a negative value of 
the coupling between the DM and Higgs boson is needed 
in the larger DM mass region 
so that the cancellation 
between $SS\to h^{\ast}\to WW^{\ast}$ 
and the four-points interaction $SSWW$ can occur 
to satisfy the relic abundance.
However, this enhances $\sigsip$ to be excluded by the XENON1T measurements. 
We also note that, due to the OPAL exclusion, 
a smaller DM mass region results in a larger $\Delta^\pm$ value. 
For the co-annihilation scenario, one can see that $\Delta^\pm>23\gev$ 
as shown in the right panel of Fig.~\ref{fig:fig2}.

\item[]\textbf{Region (d)},
it is totally opposite to the region \textbf{(a)}. 
The relic density at this region 
mainly comes from $SS\to W^\pm W^{\mp(\ast)}$ annihilation. 
Therefore, $\Delta^0$ shall be large enough  
to suppress co-annihilation and its lower bound is varied with respect to $m_S$.
The mixed scenario can reach a larger DM mass region 
as compared with the co-annihilation scenario. On the other hand,  
the lower limit of $\Delta^\pm$ for the mixed scenario from the OPAL exclusion is more released.
As shown in the right panel of Fig.~\ref{fig:fig2}, 
the mass splitting $\Delta^\pm \gsim 19$ GeV 
for the mixed scenario.

\item[]\textbf{Region (e)},
the LEP-II exclusion is presented. 
Together with the current XENON1T constraint, 
they yield two important upper limits: 
$\Delta^0<8.8\gev$ and $m_S<72\gev$.
\end{itemize}

As mentioned in the previous section, the current searches at the LHC 
is not yet to bite the parameter space.
However, we find that the extended searches at the LHC Run 3,
especially for the compressed mass spectra searches, 
can probe both co-annihilation and mixed scenarios.
On the left panel of Fig.~\ref{fig:fig2}, 
we show the future prospect contours of 2$\sigma$ significance 
from the LHC compressed mass spectra searches 
with the integrated luminosity of 250 ${\rm fb}^{-1}$ (black line), 
500 ${\rm fb}^{-1}$ (green line),  
and 750 ${\rm fb}^{-1}$ (orange line). 
Here, we fix $\lambda_S=0$, $\lambda_2=1$ and $\Delta^{\pm} = 28\gev$ 
as a benchmark point and only 
recast the dimuons final state of \textbf{SR-E-low} signal
region with the muons invariant mass window: 3.2 GeV $\leq m_{\mu^+\mu^-} \leq$ 5 GeV~\cite{Aad:2019qnd}.
The significance is given by $z = S/\sqrt{B+\delta_B^2}$,
where $S$ is the number of signal event, 
$B$ is the number of background event
and $\delta_B$ is the background uncertainty 
which is assumed to be $10\%$ of the number of background event at the future LHC.
One can see that both scenarios 
can be partly probed  
at the LHC Run 3 with ${\cal{L} }$ = 250 ${\rm fb}^{-1}$. 
The parameter space in the co-annihilation scenario can be mostly covered
if the integrated luminosity reach 500 ${\rm fb}^{-1}$, 
while one needs the luminosity about 750 ${\rm fb}^{-1}$ to probe the whole
parameter space in the mixed scenario.
A combined analysis of various signal regions, as the strategy presented in Ref.~\cite{Aad:2019qnd},
will certainly improve the significance, however, 
it is beyond the scope of this work. 
On the other hand, the mass-splitting $\Delta^{\pm}$ is sensitive to the stransverse mass $m_{T2}$, 
similar recasting method can be done according to Ref.~\cite{Aad:2019qnd}.
We will return to these two parts in a future work.

\begin{figure}[htbp]
\begin{centering}
\includegraphics[width=0.45\textwidth]{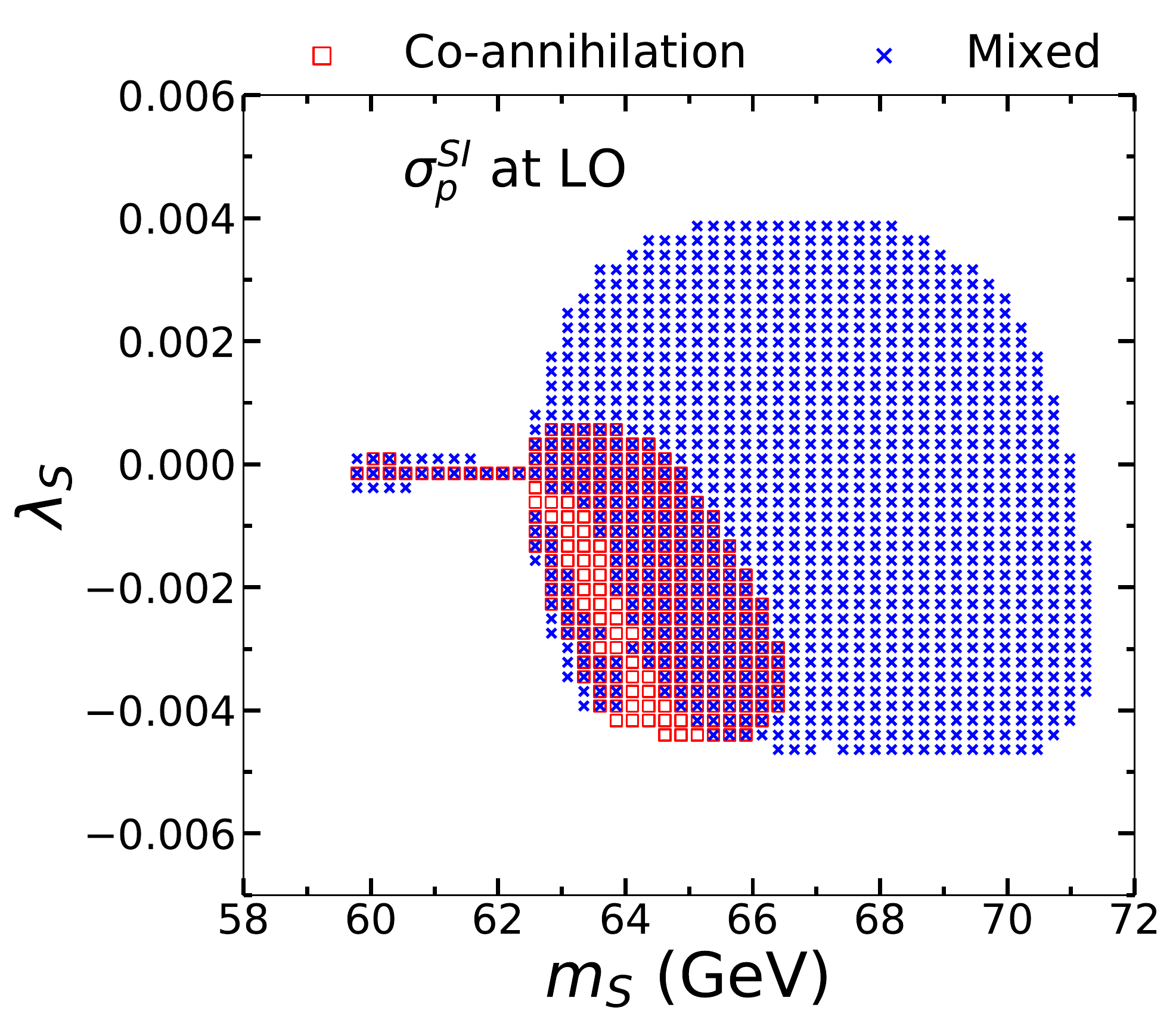}
\includegraphics[width=0.45\textwidth]{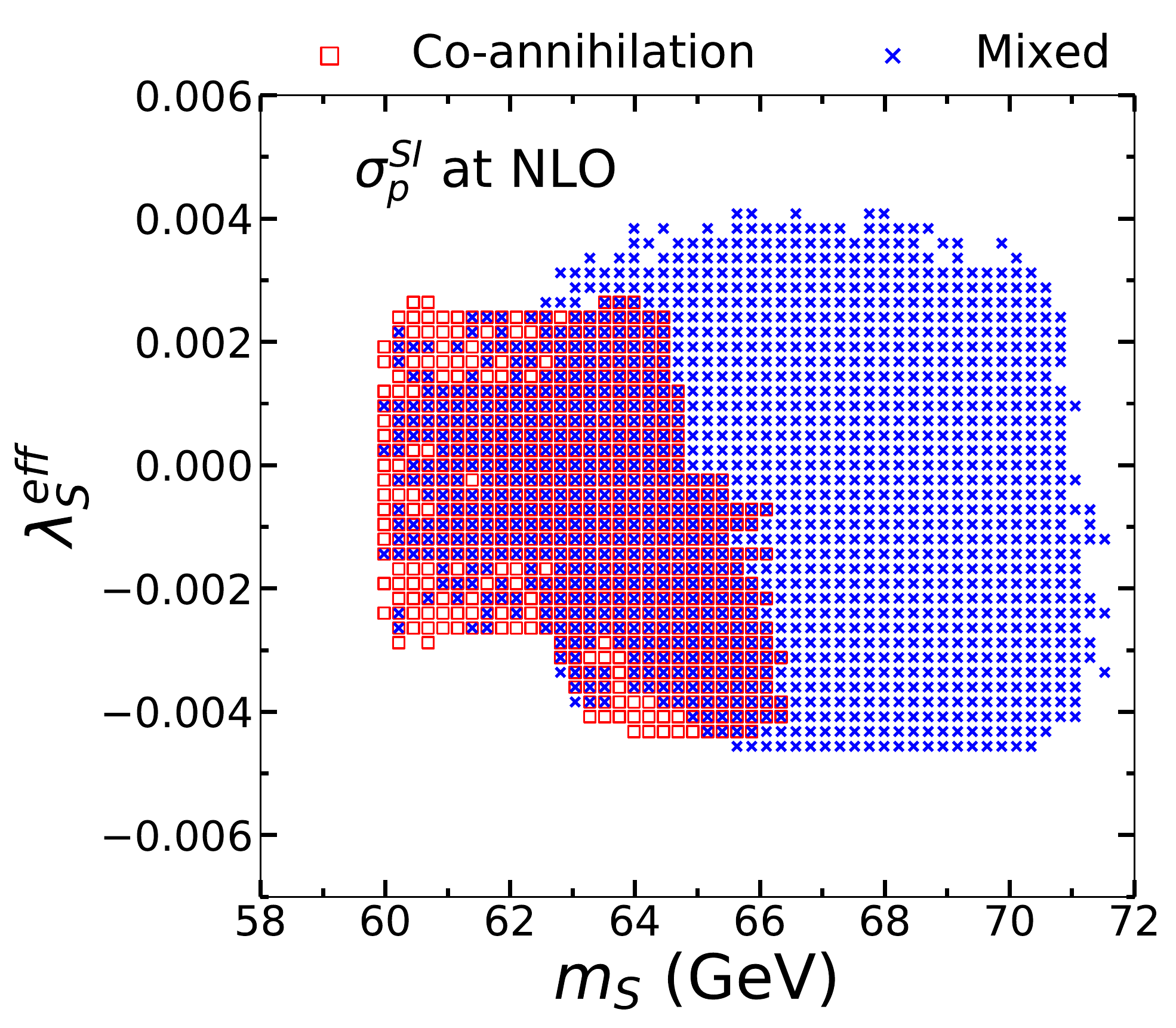}
\includegraphics[width=0.45\textwidth]{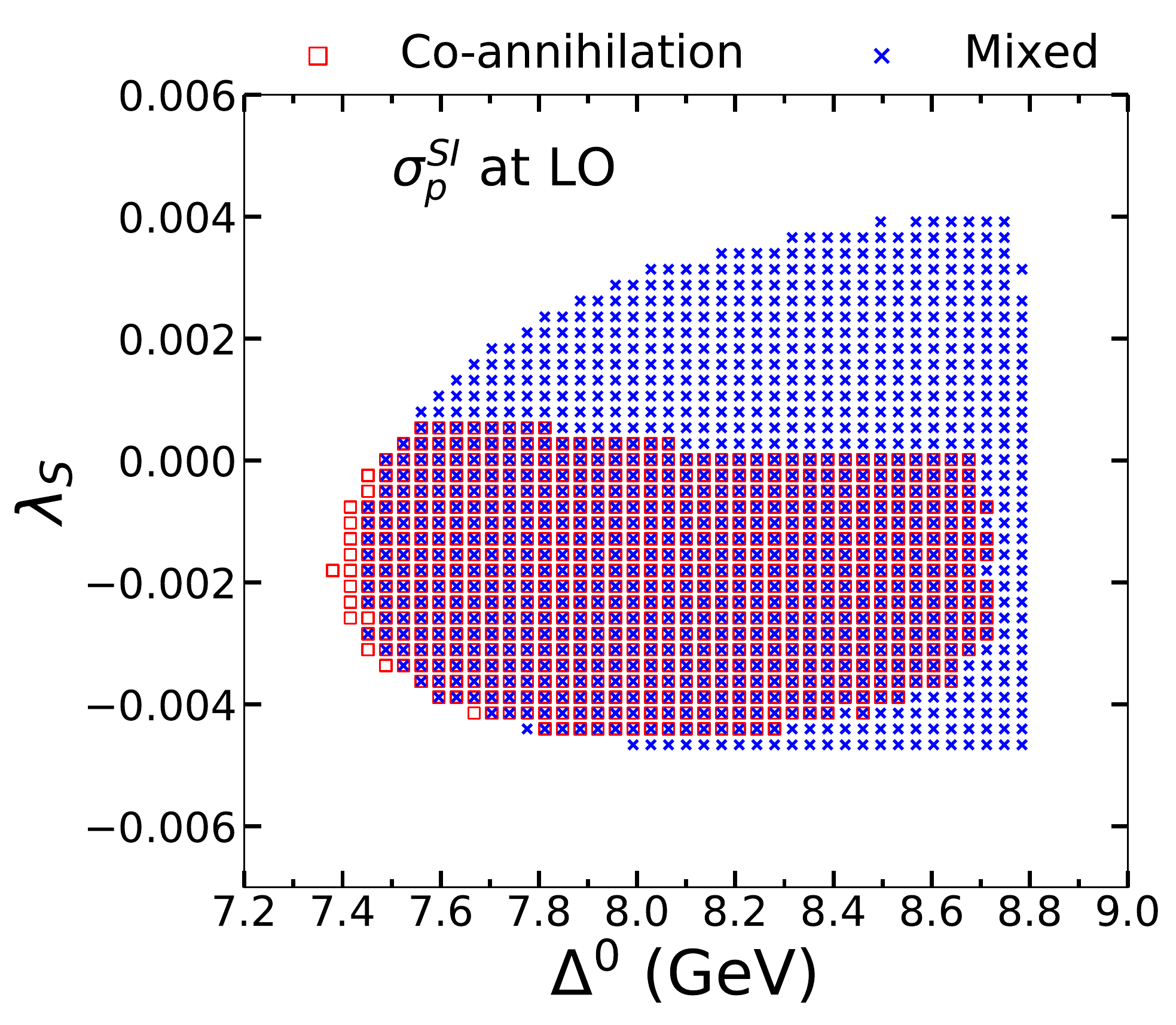}
\includegraphics[width=0.45\textwidth]{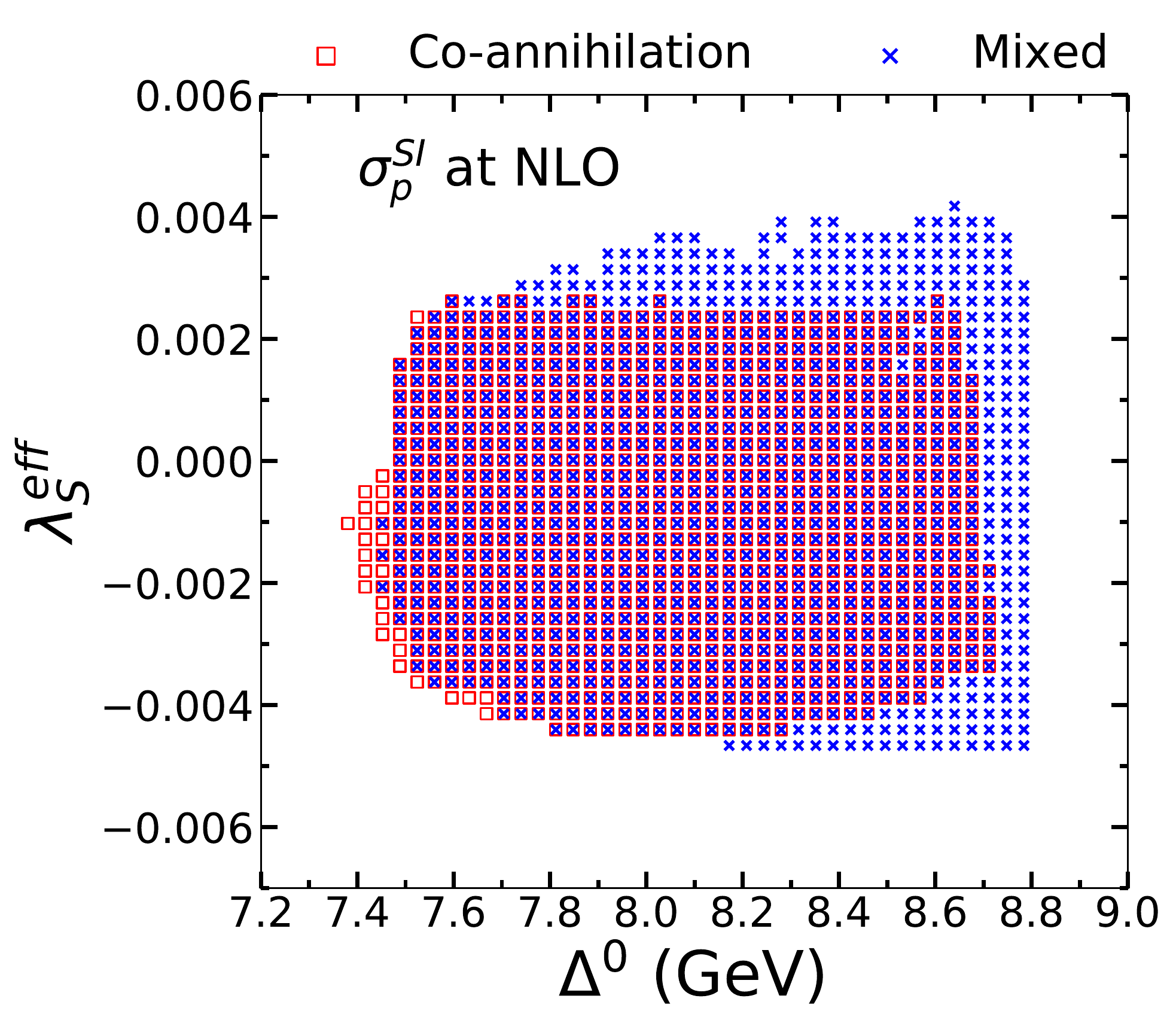}
\caption{The scatter plots in $2\sigma$ allowed region.
Left panels indicate the $\sigsip$ computed at LO 
while right panels represent the $\sigsip$ computed at NLO. 
Note that the effective coupling $\lambda_S^{\rm eff} = \lambda_S + \delta \lambda$ in the right panels.
The color scheme is the same as Fig.~\ref{fig:relic-lowmass}.} 
\label{fig:basic_plane}
\end{centering}
\end{figure}

As shown in Eq.~\eqref{eq:NLORatio}, the NLO effects on the DM-proton 
scattering cross section can be understood 
by comparing the $\lambda_S$ and $\lambda_S^{\rm eff}$ couplings. 
In Fig.~\ref{fig:basic_plane}, we show the $2\sigma$ allowed region taking into account all constraints. 
The elastic scattering cross section $\sigsip$ in the two left panels are based on 
the tree level coupling $\lambda_S$, and the two right panels are based on $\lambda_S^{\rm eff}$. 
Let us start with the two upper panels of Fig.~\ref{fig:basic_plane}. 
We can see two interesting regions: i) $m_S \lesssim m_h/2$, 
and ii) $m_S \gtrsim m_h/2$.  
Because of the Higgs resonance or co-annihilation process in the early universe, 
the $\lambda_S$ of the first region is required to be small to fulfill the relic density constraint, however 
such a small coupling makes an undetectable $\sigsip$ in the present XENON1T experiment.   
Particularly, the $\lambda_S$ in the co-annihilation region can be even smaller than 
the one in the annihilation (Higgs resonance) region. 
However, the NLO corrections can enhance $\sigsip$ to be more detectable 
in the near future direct detection experiments, 
and thus the effective coupling $|\lambda_S^{\rm eff}|$ can reach about $0.003$ as seen in the upper right panel.   
Note that $\lambda_S^{\rm eff}$ can be generated by gauge boson loops 
as the $g^2g^{\prime 2}$ term shown in Eq.~\eqref{eq:l4run} from one-loop RGEs. 
This reveals that some fine-tuning of the parameters is needed in order to reach 
a tiny $|\lambda_S^{\rm eff}|$ which is much less than $\mathcal{O}(10^{-3})$ in Fig.~\ref{fig:basic_plane}.

Regarding the second region $m_S \gtrsim m_h/2$, the four-vertex diagram $SSWW$ process becomes too sufficient 
to reduce the relic density. 
Hence, 
$\lambda_S$ needs to be negative so that a cancellation between 
this diagram and $SS\to h^{\ast}\to WW^{\ast}$ can occur. 
We can see that the exact cancellation happens
at the strip region, $\lambda_S\approx -(s-m_h^2)/(2v^2)$,
where the mixed scenario is absent because the universe is over abundant. 
Once the co-annihilation mechanism is triggered, the correct relic density can be still obtained in this region and nearby. 
For co-annihilation scenario, the above cancellation also makes an upper bound of DM mass with respect to 
the size of $\lambda_S$.

\begin{figure}[htbp]
\begin{centering}
\includegraphics[width=0.65\textwidth]{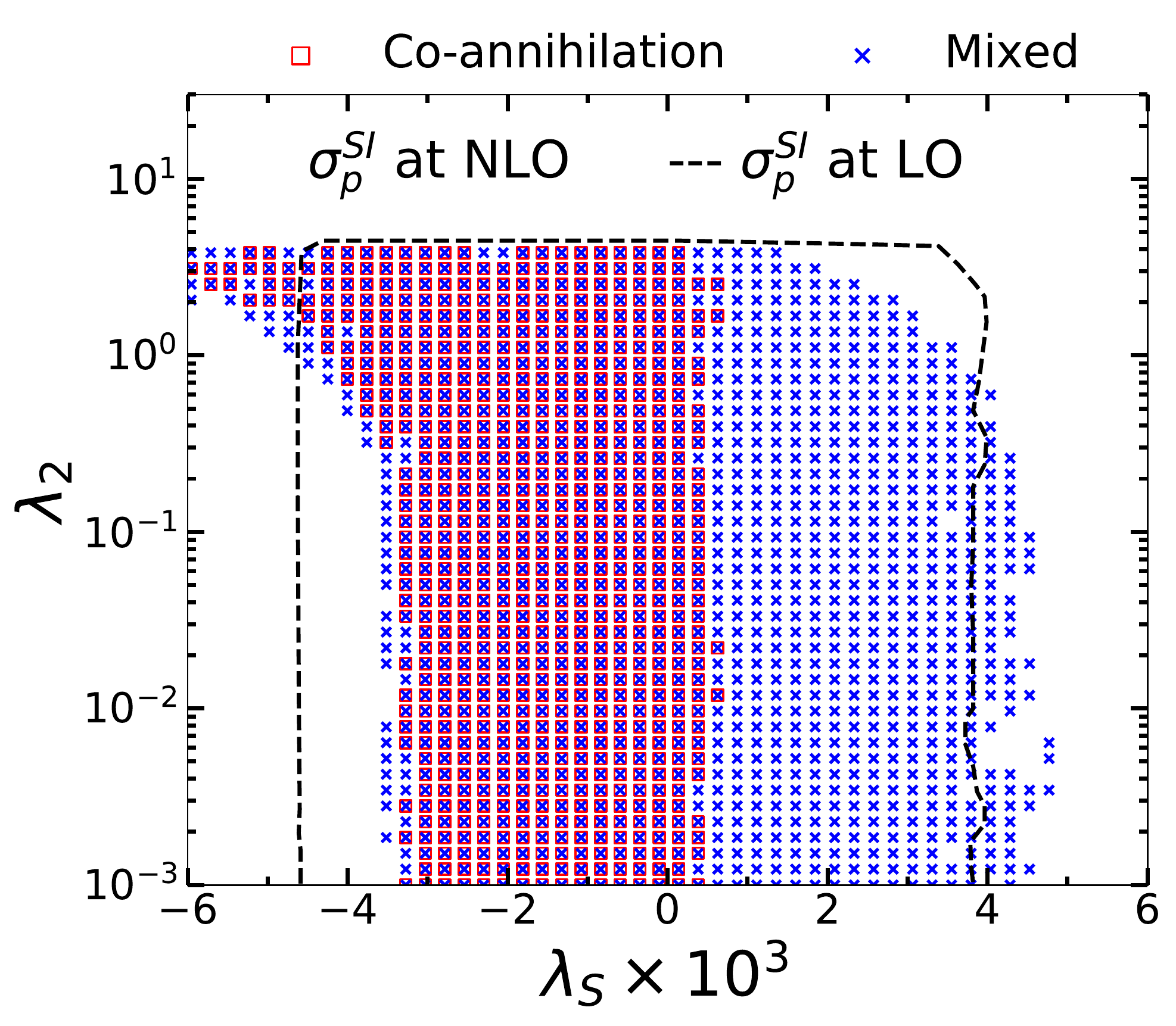}
\caption{The scatter plots in $2\sigma$ allowed region on the plane of ($\lambda_S \times 10^3$, $\lambda_2$). 
The scattered points represent $\sigsip$ at NLO while 
the region inside black dashed contours indicates that $\sigsip$ is computed at LO. 
}
\label{fig:loop_effect0}
\end{centering}
\end{figure}

The two lower panels of Fig.~\ref{fig:basic_plane} show the $2\sigma$ distribution of 
$\lambda_S$ (left panel) and $\lambda_S^{\rm eff}$ (right panel) as a function of $\Delta^0$.  
The maximum value of $\Delta^0$ for co-annihilation scenario is due to the combined constraints 
from LEP-II and relic density.   
Because a large positive value of $\lambda_S$ in this scenario is disfavored 
when the $SS\to WW^*$ annihilation process kinematically opens, 
the asymmetry between positive and negative $\lambda_S$ can be found in the co-annihilation scenario.   
In addition, the co-annihilation gradually losses the power if splitting $\Delta^0$ increases 
and therefore the positive $\lambda_S$ has to be decreased once co-annihilation dominates at the early universe. 
However, the negative $\lambda_S$ is rather favorable because of the cancellation between diagrams of 
the four-points interaction and $SS\to h^{\ast}\to WW^{\ast}$. 
Again, we can see from the lower-right panel that the NLO corrections can generally increase 
the size of $|\lambda_S^{\rm eff}|$, 
especially for the positive $\lambda_S$ region which has been excluded if only tree level is considered. 
Note that the value of $|\lambda_S^{\rm eff}|$ can be smaller than $|\lambda_S|$ 
if the loop correction parameter $\delta \lambda$ and $\lambda_S$ are  opposite signs 
as shown in Appendix~\ref{app:supp}.

In Fig.~\ref{fig:loop_effect0}, the NLO effects are more distinguishable on the plane of ($\lambda_S \times 10^3$, $\lambda_2$). 
The coupling $\lambda_2$ is not involved in LO processes, 
but it causes some interesting behaviors once the NLO corrections are included.   
There are two main features. 
The first feature is that some cancellations between the tree level diagrams 
and loop level diagrams are taking place. 
Particularly, these tree-loop cancellations can be found 
in the region of $\lambda_S \lsim -4.8\times 10^{-4}$ and $\lambda_2>1$ 
as well as the region $\lambda_S>4\times 10^{-4}$
where the parameter space were excluded from LO computation but they can be saved when including NLO corrections.
The second feature is that the NLO contribution generally increases $\sigsip$.  
Apart from the tree-loop cancellation region, 
more parameter space have been ruled out comparing with only LO computation.

\begin{figure}[htbp]
\begin{centering}
\includegraphics[width=0.45\textwidth]{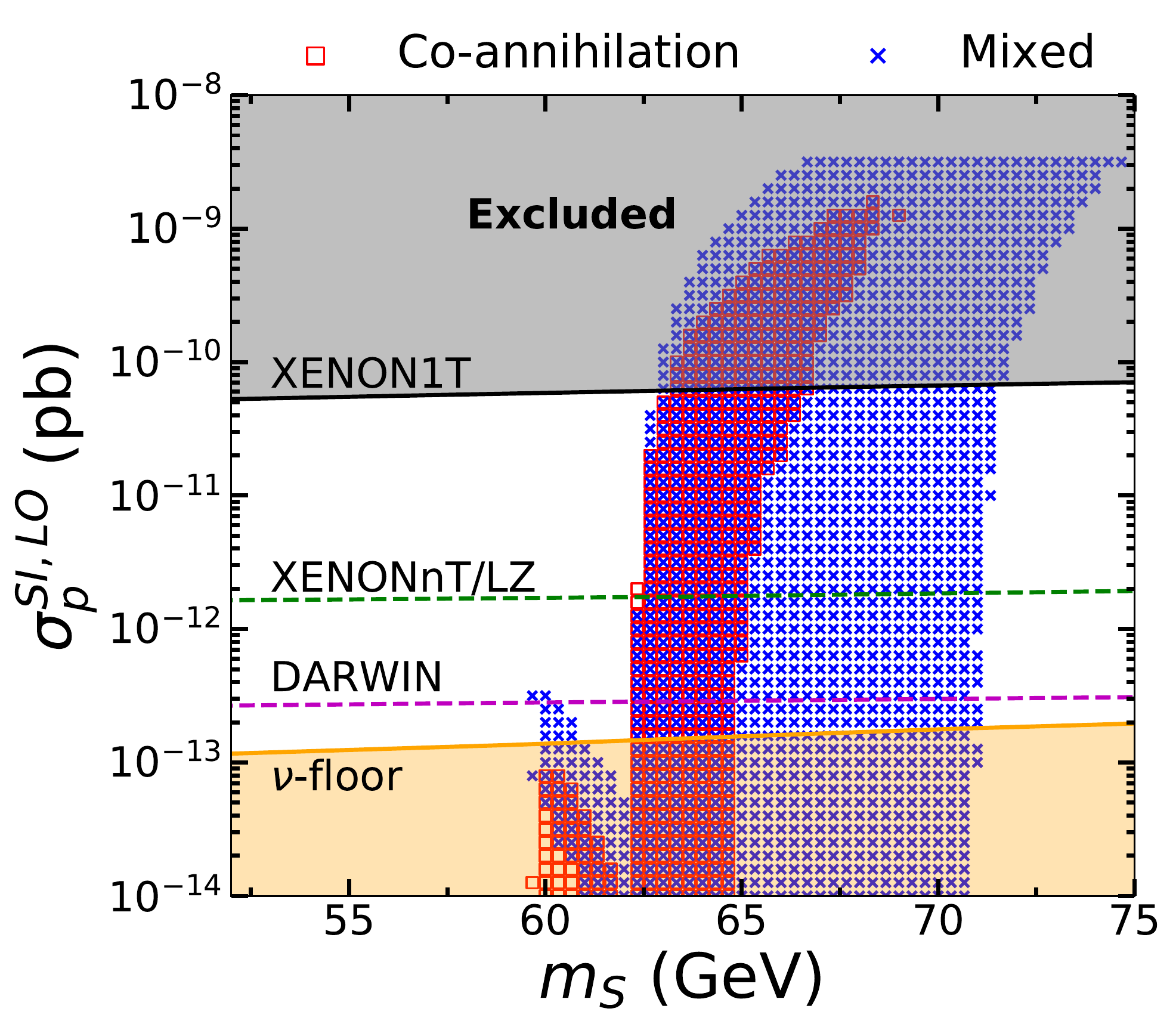}
\includegraphics[width=0.45\textwidth]{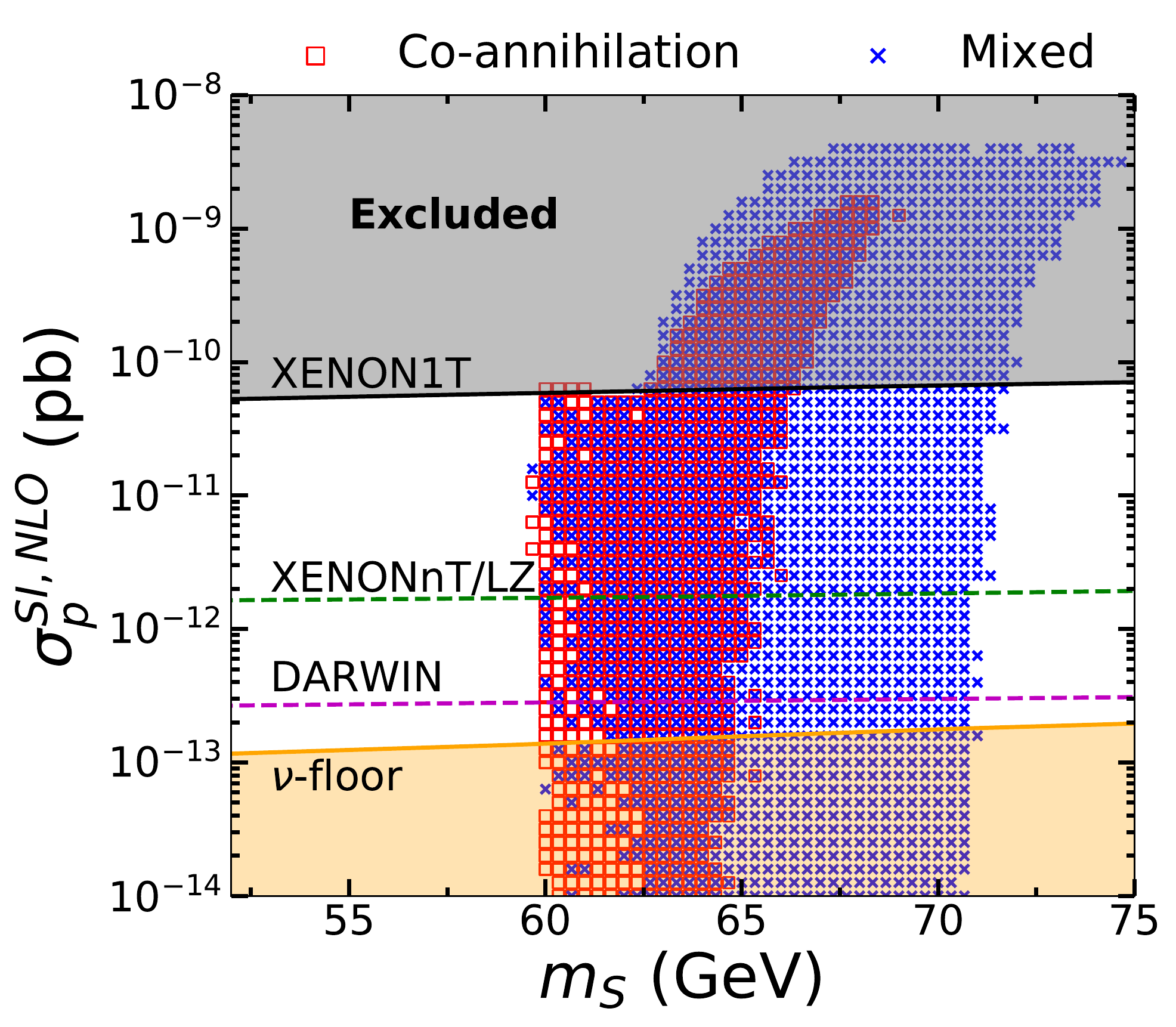}
\caption{
The scatter plots on the ($m_S$, $\sigma^{\rm SI}_p$) plane. 
Allowed points are in $2\sigma$ agreement with all constrains except
the XENON1T measurements.   
The DM-proton scattering cross section is computed at LO in the left panel 
while the one taking into account the one loop correction is presented in the right panel.
The color scheme for the scatter points is the same as Fig.~\ref{fig:relic-lowmass}. 
The solid black line represents the limit from XENON1T~\cite{Aprile:2018dbl}.  
The projected sensitivities of LZ~\cite{Akerib:2015cja} and DARWIN~\cite{Aalbers:2016jon} 
are dashed green and dashed magenta lines, respectively.  
The orange region is the neutrino background \cite{Billard:2013qya}.
}
\label{fig:sigsip}
\end{centering}
\end{figure}

Fig.~\ref{fig:sigsip} shows the DM-proton scattering cross section as a function of DM mass, $m_S$, 
at tree-level (left panel) and one-loop level (right panel).
Note that the DM direct detection constraints are not included in 
the scatter point regions to demonstrate its exclusion power. 
Again, the inelastic scattering can be neglected at the region 
$\Delta^0\gtrsim\mathcal{O}(200\kev)$~\cite{Arina:2009um}.  
Hence, the only significant contribution to the detection rate is 
the elastic scattering via $t$-channel Higgs exchange.
 
Because both scenarios are generally required a small $\lambda_S$ to fulfill the relic density constraints, 
making a drawback to detect the co-annihilation and Higgs resonance region. 
In the left panel, we can see that the elastic scattering cross section in  
the resonance region, $m_S \lsim m_h/2$, 
is overall lower than the future projected sensitivities from 
LZ~\cite{Akerib:2015cja} and DARWIN~\cite{Aalbers:2016jon}. 
The co-annihilation region at $m_S<m_h/2$ is even below the neutrino floor and 
hard to be detected under current strategies of DM direct detection.

Strikingly, once the next-leading order correction for $\sigsip$ is considered, 
the elastic scattering cross section $\sigsip$ in the Higgs resonance
and the co-annihilation regions are both significantly enhanced. 
Thanks to the loop contributions, especially from the large $|\lambda_2|$,
co-annihilation region can be testable in the future direct detection searches.

\begin{figure}[htbp]
\begin{centering}
\includegraphics[width=0.6\textwidth]{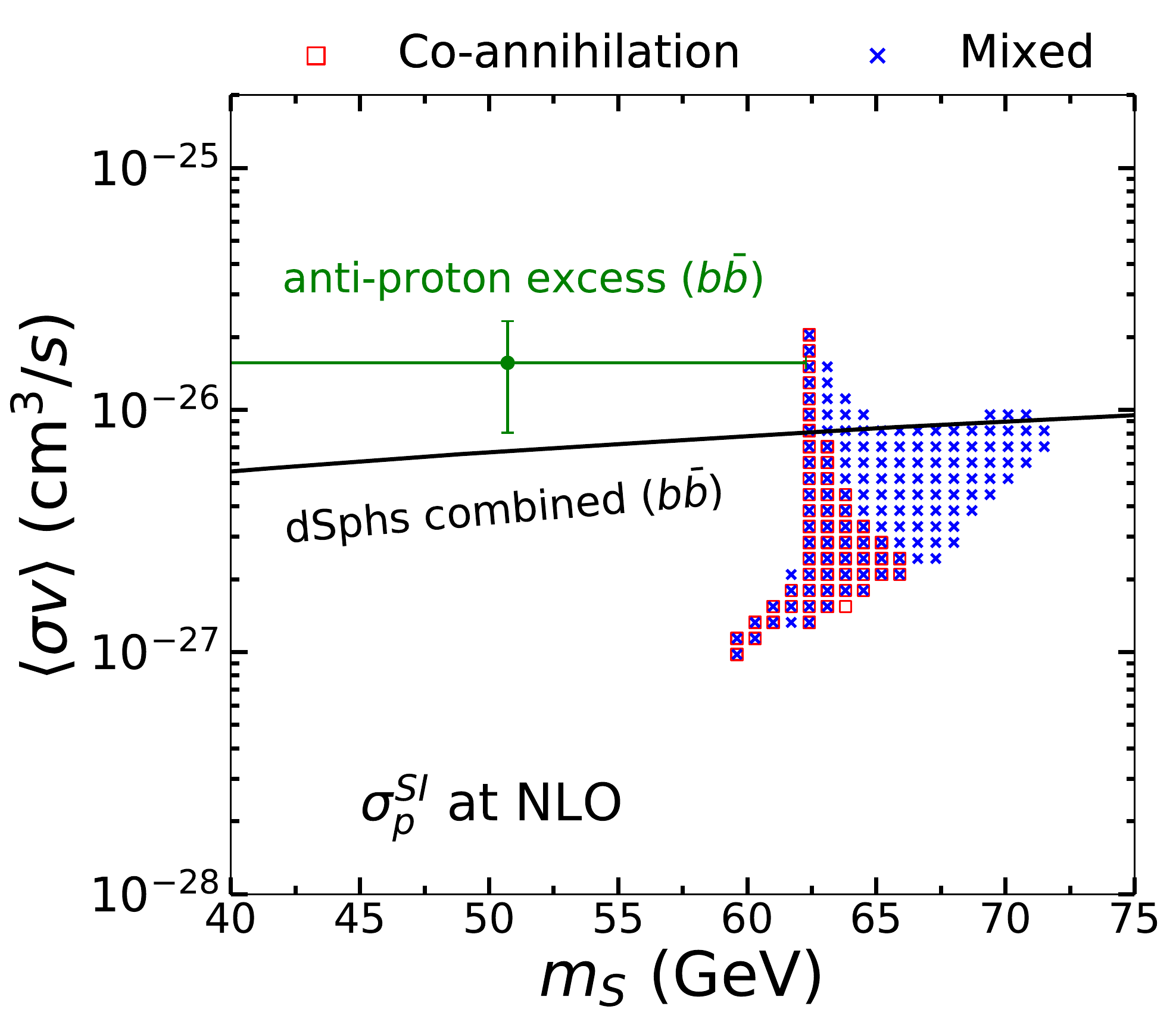}
\caption{
The scatter plots in $2\sigma$ allowed region on the ($m_S$, $\langle \sigma v \rangle$) plane. 
The color scheme for the scatter points is the same as Fig.~\ref{fig:relic-lowmass}. 
The solid black line represents the combined limit for DM annihilating into $b \bar{b}$ 
from observations of dSphs by Fermi-LAT, HAWC, HESS, MAGIC and VERITAS \cite{Oakes:2019ywx}.
The green error bar is the $1\sigma$ signal region for the antiproton excess~\cite{Cui:2018nlm}.
}
\label{fig:sigmav}
\end{centering}
\end{figure}

Lastly, we would like to discuss the detection of the DM annihilation at the present. 
In Fig.~\ref{fig:sigmav}, we show the $2\sigma$ distribution in ($m_S$, $\langle \sigma v \rangle$) plane, 
allowed by all constraints in Table~\ref{tab:likelihood}. 
With a conservative treatment, we have taken only the Fermi 15 dSphs gamma ray data into our total likelihood. 
However, we also show the current most stringent limit (solid black line) 
which is obtained by combing the latest data 
from Fermi-LAT, HAWC, HESS, MAGIC and VERITAS~\cite{Oakes:2019ywx}. 
To illustrate the antiproton anomaly, we also present the signal region 
for $b\bar{b}$ final state~\cite{Cui:2018nlm} as a comparison.

For the region of $m_S < m_h/2$, the dominant annihilation channel is $SS\to h \to b\bar{b}$  
where $\lambda_S$ is suppressed due to the relic density constraint.
Thus, the velocity averaged cross section at this region is a relatively small value, 
$\sigmav$ $\sim 10^{-27}~{\rm cm}^3 s^{-1}$. 
The Higgs resonance presents at the region of $m_S\simeq m_h/2$,
but the large cross section part can be excluded 
if one takes into account the combined limit (black line).
The DM annihilation at the present universe with mass $m_S>m_h/2$ are almost entirely to $WW^*$ final state 
and its spectrum $dN/dE$ is similar to the $b\bar{b}$ final state. 
Except a small part of blue crosses (mixed scenario), most of parameter space 
in this region are still survived. 
Interestingly, the viable regions of the model including 
the mixed and co-annihilation scenarios are in agreement 
with the antiproton anomaly within $3\sigma$~\cite{Cui:2018nlm}.
Hence, if the antiproton anomaly would be confirmed in the future, 
one would expect to see a signal at this DM mass region from 
the future HL-LHC compressed mass spectrum searches.


\section{Summary and future prospect\label{sec:conclusion}}


The compressed mass spectrum in the i2HDM is an interesting topic. 
It can escape from most of current collider constraints and  
its relic density in this region can be mainly governed by 
the $S$ and $A$ co-annihilations. 
To obtain the correct relic density and minimize the contribution from annihilation channels, 
there are two main features of the co-annihilation scenario. 
The first one is the requirement for a negligible $|\lambda_S|$ near the Higgs resonance region. 
As a type of Higgs portal DM model, 
one can intuitively reduce $\lambda_S$ to minimize the annihilation contributions.  
However, this method only works at the region where $SS\to W^+ W^{-\ast}$ is not open. 
Once this channel is kinematically-allowed at the early universe, 
the $SSW^+W^-$ four-points interaction 
is very sufficient to reduce the relic density.
To lower down the contribution from the $SSW^+W^-$ four-points interaction, 
one has to choose a negative sign of $\lambda_S$ to make a cancellation among 
four-points interaction and $s$-channel Higgs exchange. 
Therefore, the second feature is that a negative $\lambda_S$ is needed 
if $SS\to W^+ W^{-\ast}$ channel is kinematically-allowed at the early universe.

Because of a small or negative $\lambda_S$,
the NLO corrections for DM-nucleon elastic scattering become more important.
We first fix the input scales of our scan parameters at the EW scale,
the effective coupling $\lambda_S^{\rm eff}$ will be modified from quantum corrections 
at the low energy DM direct detection scale.
At the tree-level, a small value of $|\lambda_S|$ predicts 
a cross section $\sigsip$ smaller than the neutrino floor. 
However, it can be testable if $\sigsip$ includes the NLO corrections.
If the value of $|\lambda_S|$ is large and $\lambda_S$ is negative, 
a cancellation between LO and NLO contribution may be needed   
in order to escape from the present XENON1T constraint. 
Such a NLO contribution is sensitive to not only 
the mass splitting $\Delta^\pm$ but also the coupling $\lambda_2$. 
Interestingly, the coupling $\lambda_2$ plays no role at the tree-level phenomenology, 
neither Higgs nor DM. 

Motivated by the non-trivial correlations between compressed mass spectra, co-annihilations, 
and NLO corrections of $\sigsip$, we conduct a global scan to comprehensively explore 
the parameter space of compressed mass spectra in the i2HDM, including five parameters 
($m_S$, $\Delta^0$, $\Delta^\pm$, $\lambda_2$, $\lambda_S$) at the EW scale. 
Particularly, the parameters $\Delta^0$ and $\Delta^\pm$ are adopted 
for searching the compressed mass spectrum and co-annihilation scenario. 
By using the profile likelihood method, 
the survived parameter space were subjected to constraints from 
the theoretical conditions (perturbativity, stability, and tree-level unitarity), 
the collider limits (electroweak precision tests, LEP, and LHC),  
the relic density as measured by PLANCK, the $\sigsip$ limit from XENON1T, 
and the $\sigmav$ limit from Fermi dSphs gamma ray data. 
For the computation of $\sigsip$, the LO and NLO calculations are considered separately in the likelihood. 
We have shown the $2\sigma$ allowed points grouped by co-annihilation scenario ($f_{ann} < 15\%$) and 
mixed scenario ($f_{ann} > 15\%$) in two dimensional projections of the parameter space. 

We found that the viable parameter spaces for the co-annihilation scenario 
are located at the $60\gev\lesssim m_S \lesssim 66\gev$, 
$7.4\gev\lesssim \Delta^0 \lesssim 8.7\gev$, and $\Delta^\pm\gtrsim 23.0\gev$.
The NLO correction $\delta\lambda$ is sensitive to $\lambda_2$ when it is greater than 
0.2
while the contribution from $\Delta^\pm$ dominates the $\delta\lambda$ if $\lambda_2<0.2$.
Due to the interplay between OPAL exclusion and PLANCK relic density constraint,
the co-annihilation scenario cannot be realized with the condition 
$-10^{-3}\lesssim\delta\lambda \lesssim 0.0$ when the contribution from $\Delta^\pm$ 
dominates the $\delta\lambda$. 

Next, the correlation between $\lambda_S$ and $\lambda_2$ is non-trivial at the NLO level. 
For $\lambda_2>1$, the same sign between $\lambda_S$ and $\delta\lambda$ can enhance $\sigsip$ so that 
the parameter space with a large value of $\lambda_S$ and $\lambda_2$ is excluded. 
For $\lambda_2<1$, the $\Delta^\pm$ plays the role to NLO corrections. 
Therefore, the cross section $\sigsip$ for co-annihilation scenario can be significantly enhanced  
but ruled out by XENON1T. It becomes testable in the future direct detection experiments.

Finally, 
the $95\%$ allowed region predicted in the model coincides with the AMS-02 antiproton anomaly 
within $3\sigma$ range. This region can be also tested by  
the compressed mass spectra searches at the LHC.  
Although we found that the region is not sensitive to the current LHC searches, 
it can be partially probed with future  
luminosity 250 ${\rm fb}^{-1}$ and mostly probed with 
luminosity 750 ${\rm fb}^{-1}$ as shown in the left panel of Fig.~\ref{fig:fig2}.

\newpage

\section*{Acknowledgments}

We thank Tomohiro Abe and Ryosuke Sato for some useful discussions and suggestions 
in the part of spin-independent cross section at the next leading order in i2HDM.
Y.-L. S. Tsai was funded in part by the Chinese Academy of Sciences Taiwan Young
Talent Programme under Grant No. 2018TW2JA0005.
V.Q. Tran was funded in part  
by the National Natural Science Foundation of China under Grant Nos.\ 
11775109 and U1738134. 

\appendix

\section{Spin-independent cross section at the next leading order}
\label{app:sigsip_NLO}
\vspace{-0.5em}

\allowdisplaybreaks
In this Appendix, we outline the calculations of spin-independent cross section at the next leading order from Ref.~\cite{Abe:2015rja}. We have checked the consistency of our numerical results. First, the effective interaction of the dark matter and quark/gluon can be represented as
\begin{align}
 {\cal L}_{\rm eff.}
=&
\frac{1}{2} \sum_{q=u,d,s} \Gamma^{q} S^2 (m_q \bar{q}q)
- \frac{1}{2} \frac{\alpha_s}{4 \pi} \Gamma^{G} S^2 G^{a}_{\mu \nu} G^{a \mu \nu}
\nonumber\\
&
+ \frac{1}{2 m_S^2} \sum_{q = u,d,s,c,b} \left[
      (\partial^{\mu}S) (\partial^{\nu}S) \Gamma^q_{\text{t2}} {\cal O}^{q}_{\mu \nu}
      - S (\partial^{\mu} \partial^{\nu}S) \Gamma'^q_{\text{t2}} {\cal O}^{q}_{\mu \nu}
\right],
\label{eq:eff_int}
\end{align}
where ${\cal O}^q_{\mu\nu}$ is the quark twist-2 operator with the following form,
\begin{align}
{\cal O}^{q}_{\mu \nu}
\equiv
 \frac{i}{2} \bar q \left( \partial_\mu \partial_\nu + \partial_\nu \partial_\mu - \frac{1}{2}g_{\mu\nu} \slashed{\partial} \right) q.
\end{align}
The higher twist gluon operators have been neglected in the above effective Lagrangian.
Based on Eq.~(\ref{eq:eff_int}), the scattering amplitude and spin-independent cross section of dark matter and nucleon can be written as,
\begin{align}
 i {\cal M}
=&
i m_N \left[
     \sum_q \Gamma^q f_q
     + \frac{2}{9} \Gamma^G f_g
     + \frac{3}{4} \sum_q (\Gamma^q_{\text{t2}} + {\Gamma'}^q_{\text{t2}} )(q(2)+\bar q(2) )
\right],
\label{eq:M-for-xsecSI}\\
\sigma_{\text{SI}} =& \frac{\mu^2}{4\pi m_S^2} |{\cal M}|^2,
\end{align}
where $\mu$ is the reduced mass with the form $\mu\equiv m_S m_N / (m_S + m_N)$. $m_S$ and $m_N$ are dark matter and nucleon masses, respectively.

\begin{table}
\centering
\begin{tabular}{|c|c|c|c|c|}
\hline
 & $\Gamma^q_{\text{Box}}$ & $ \Gamma^q_{\text{t2}} + {\Gamma'}^q_{\text{t2}}$ & $\delta \Gamma_h$ & $\Gamma^G_{\text{Box}}$ \\
\hline
$W$ & $(B.5)-(B.7)$ & $(B.6)$ & $(C.9)-(C.16)$ & $(D.19)$ \\
$Z$ & $(B.2)-(B.4)$ & $(B.3)$ & $(C.1)-(C.8)$ & $(D.20)$ \\
\hline
\end{tabular}
\caption{
The relevant equation numbers to calculate Feynman diagrams with effective interaction of the dark matter and quark/gluon at the next leading order from Ref.~\cite{Abe:2015rja}.  
Here $(B.5)$ means equation $(B.5)$ in Ref.~\cite{Abe:2015rja} with the $W$ boson contribution inside the one-loop box type diagram, for example.
}
\label{tab:LoopEqN}
\end{table}

At the leading order, only $\Gamma^q$ and $\Gamma^G$ are non-zero in Eq.~(\ref{eq:M-for-xsecSI}) and can be given as,
\begin{align}
\Gamma^q = \Gamma^G = \frac{\lambda_S}{m_h^2}
\end{align}
At the next leading order, we closely follow the calculations in Ref.~\cite{Abe:2015rja} with the following classifications, 
\begin{itemize}
    \item One-loop box type diagrams      
    \item One-loop Higgs vertex correction diagrams
    \item Two-loop gluon contribution diagrams 
\end{itemize}
We list relevant equation numbers to calculate above diagrams from Ref.~\cite{Abe:2015rja} in the Table~\ref{tab:LoopEqN}.
Here $(B.5)$ means equation $(B.5)$ in Ref.~\cite{Abe:2015rja} with the $W$ boson contribution inside the one-loop box type diagram, for example.

We can further define $\Gamma^q$ and $\Gamma^G$ at the next leading order as,
\begin{align}
\Gamma^q = \frac{\delta \Gamma_h(0)}{m_h^2} + \Gamma_{\text{Box}}^q,\qquad
\Gamma^G = \frac{\delta \Gamma_h(0)}{m_h^2} + \Gamma_{\text{Box}}^g.
\end{align}
Finally, according to Eq.$(3.21)-(3.24)$ in Ref.~\cite{Abe:2015rja} and arguments therein, the one-loop correction $\delta\lambda$ are combination of functions in the Table~\ref{tab:LoopEqN} with the following form, 
\begin{align}
\delta \lambda
\equiv&
\delta \Gamma_h(0)
- \delta \Gamma_h(m_h^2)
+ \frac{m_h^2}{f_N} \left( \sum_q \Gamma^q_{\text{Box}} f_q \right)
+ \frac{2}{9} \frac{m_h^2}{f_N} \Gamma^G_{\text{Box}} f_g
+ \frac{3}{4} \frac{m_h^2}{f_N} \sum_q ( \Gamma^q_{\text{t2}} + {\Gamma'}^q_{\text{t2}} )(q(2)+\bar q(2)). \label{eq:def_delta_lambda}
\end{align}
where $f_q$, $f_g$, $q(2)$ and $\bar q(2)$ are matrix elements for nucleons and $f_N \equiv\frac{2}{9} + \frac{7}{9}\sum_q f_q$. Their exact values for proton and neutron are taken from the default values of \texttt{MicrOMEGAs}~\cite{Belanger:2018mqt}. Therefore, the loop-induced effects can be simply written as the form in Eq.(\ref{eq:NLORatio}) with four input parameters : $m_S$, $\Delta^0$, $\Delta^{\pm}$ and $\lambda_2$ inside $\delta\lambda$.

\section{Supplemental figures}
\label{app:supp}

In this Appendix, we show some plots which are useful to understand 
the details of NLO effects in the parameter space.

\begin{figure}[htbp]
\begin{centering}
\includegraphics[width=0.45\textwidth]{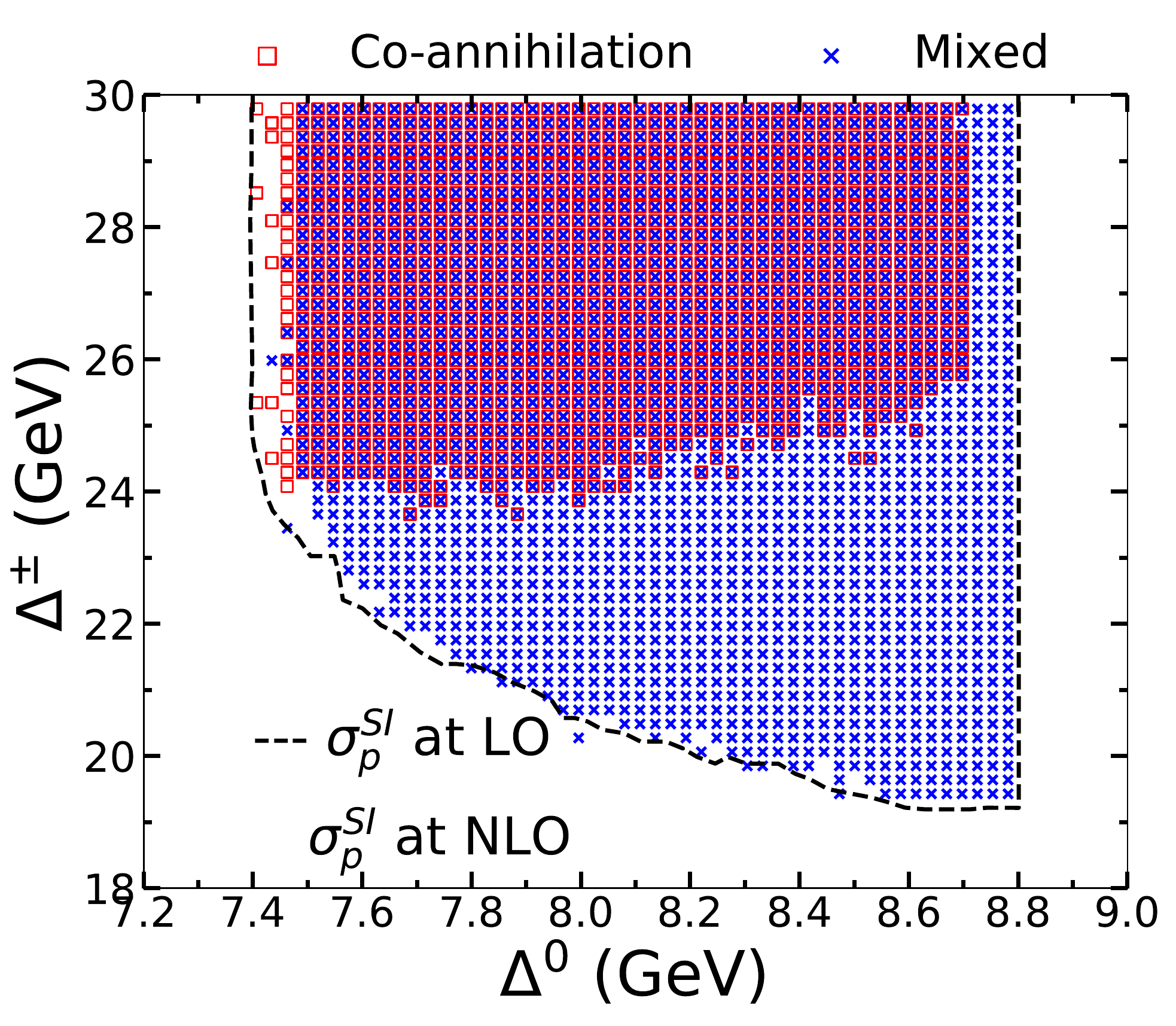}
\caption{The scatter plots in $2\sigma$ allowed region on the plane of ($\Delta^0$, $\Delta^{\pm}$).  
The scattered points represent $\sigsip$ at NLO while 
the region inside black dashed contours indicates that $\sigsip$ is computed at LO. 
The color scheme is the same as Fig.~\ref{fig:relic-lowmass}.
}
\label{fig:lneoop_effect}
\end{centering}
\end{figure}

The $2\sigma$ distribution for 
$(\Delta^0, \Delta^{\pm})$ planes 
are shown in Fig.~\ref{fig:lneoop_effect}.
The black dashed line represents $2\sigma$ contour for the LO calculation
while the scatter points correspond to the one taking into account the NLO effects.
From the left panel of Fig.~\ref{fig:lneoop_effect}, one can see that 
the NLO contribution takes effect only 
to the small $\Delta^0$ or $\Delta^{\pm}$ region, 
but it does not change the LO result significantly.

\begin{figure}[htbp]
\begin{centering}
\includegraphics[width=0.45\textwidth]{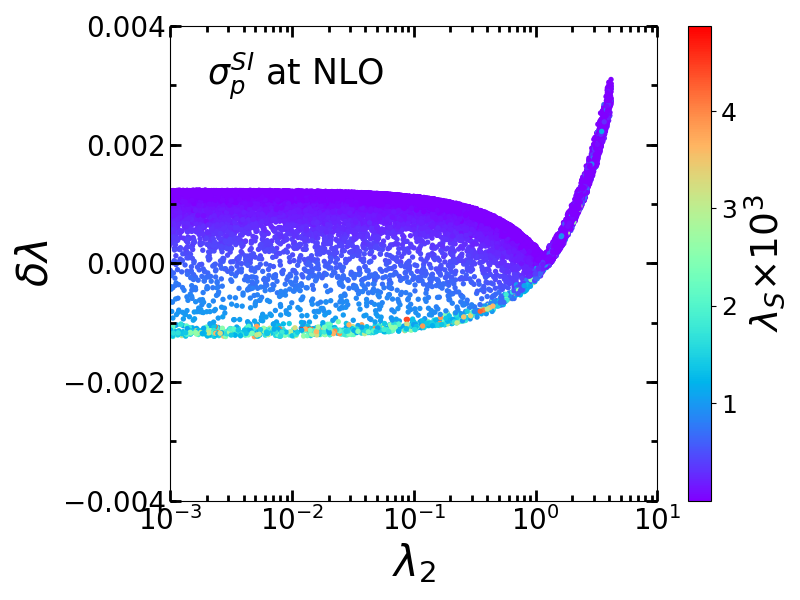}
\includegraphics[width=0.45\textwidth]{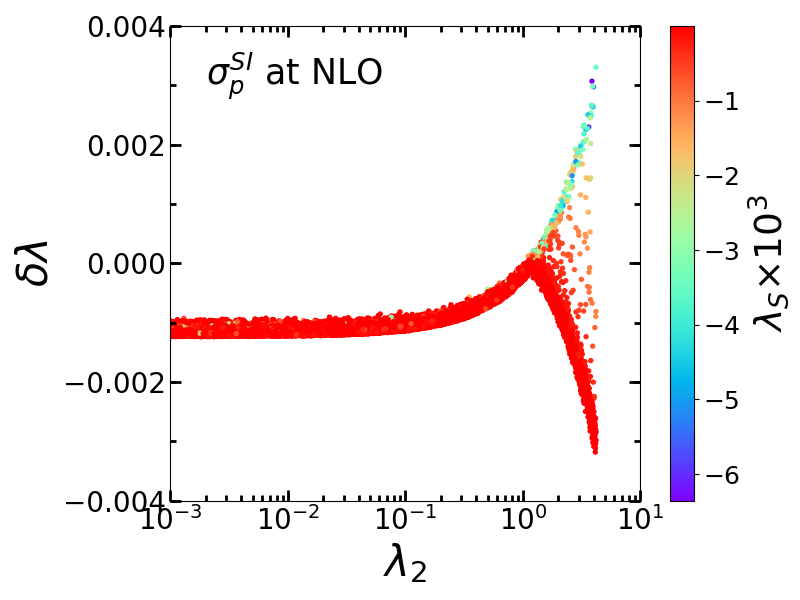}
\includegraphics[width=0.45\textwidth]{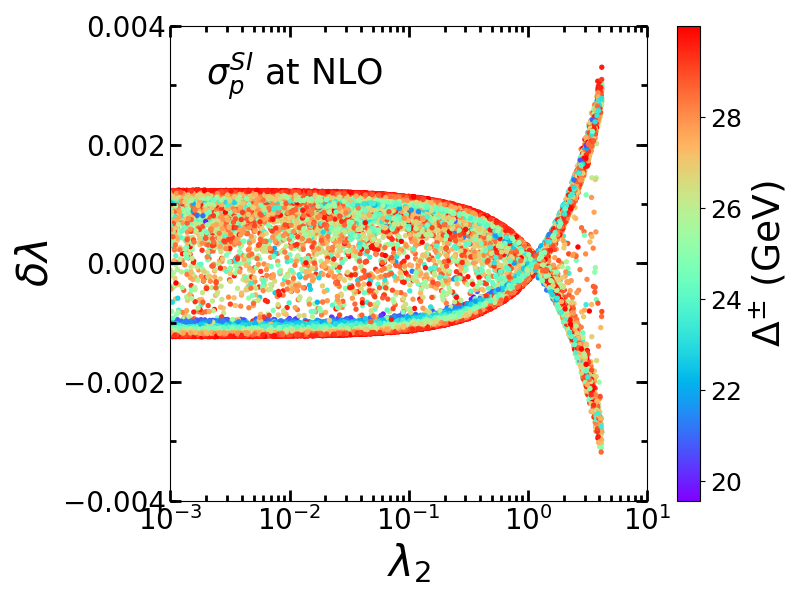}
\includegraphics[width=0.4\textwidth]{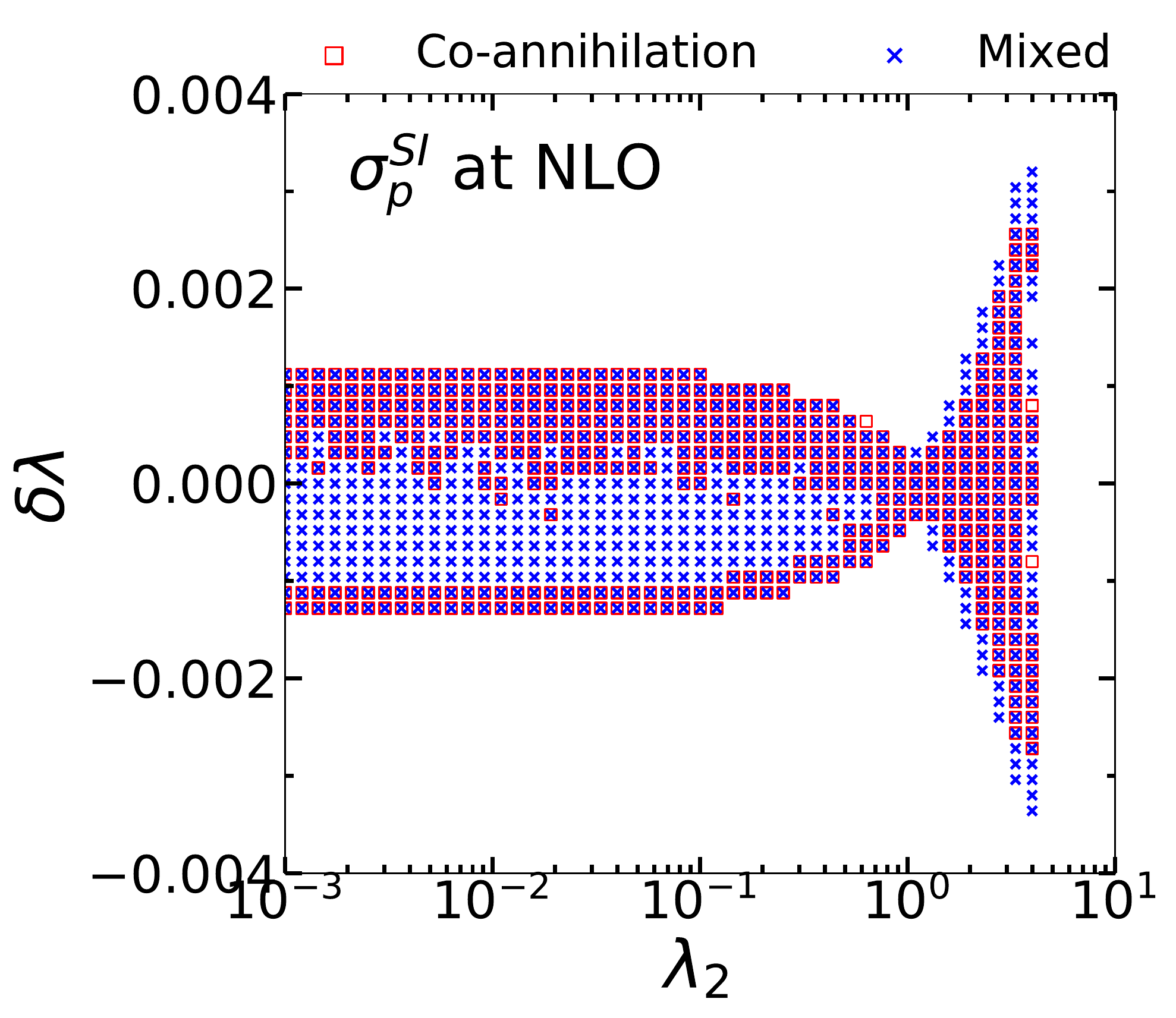}
\caption{The $2\sigma$ allowed region on the plane ($\lambda_2$, $\delta \lambda$). 
Here, $\sigsip$ is calculated at next leading order. 
}
\label{fig:loopeff}
\end{centering}
\end{figure}

In Fig.~\ref{fig:loopeff}, we discuss the loop correction parameter $\delta \lambda$ 
as a function of $\Delta^{\pm}$, $\lambda_S$, and $\lambda_2$. 
Although the $\Delta^0$ can alter the size of $\delta \lambda$, 
it is not significant due to its smallness required by the relic density constraint.
In the two upper frames of Fig.~\ref{fig:loopeff}, 
we discuss the cases of $\lambda_S>0$ (upper-left panel) and 
$\lambda_S<0$ (upper-right panel) on the ($\lambda_2$, $\delta\lambda$) plane, separately. 
The cross section $\sigsip$ is proportional to the effective coupling squared 
$(\lambda_S^{\rm eff})^2$ which is sensitive to the relative signs of 
$\lambda_S$ and $\delta\lambda$ 
if they are in the same order.
In particular, the cross section at NLO 
can even be lower than LO if the cancellation 
between $\lambda_S$ and $\delta\lambda$ occurs.
A large value of $|\lambda_S|$ tends to accompany  
with an opposite sign of $\delta\lambda$, 
while a small value of $|\lambda_S|$ can either accompany with
a same sign or opposite sign of $\delta\lambda$. 
Such a configuration results a small $\sigsip$ at NLO as well as 
escapes from the XENON1T constraint.

The scatter plots of all allowed points on the ($\lambda_2$, $\delta\lambda$) plane 
are shown in the lower frames of Fig.~\ref{fig:loopeff}. 
The color codes indicate the value of $\Delta^\pm$ (lower-left panel) and the governed relic density channels (lower-right panel).
For the region of $\lambda_2>0.2$, the loop correction parameter $\delta \lambda$ highly depends on $\lambda_2$. 
Especially, due to the cancellation between loop diagrams, 
$\delta \lambda$ 
drops down at $\lambda_2 \sim 1.4$. 
However, it significantly increases 
for a larger value of $\lambda_2$.
On the other hand, $\Delta^\pm$ plays a significant role in $\delta \lambda$ for the region of $\lambda_2 \lsim 0.2$. 
In particular, the lower limit on $\Delta^\pm$ 
which is due to the OPAL exclusion, 
can give a lower limit on the loop correction $\delta\lambda$. 
Furthermore, in this region, $|\delta\lambda|$ is restricted to be small ($|\delta\lambda| < 1.1 \times 10^{-3}$) 
due to our choice of $\Delta^\pm < $ 30 GeV.
As aforementioned, the OPAL exclusion is more stringent  
on the co-annihilation than the mixed scenario, 
hence a stronger lower limit on $|\delta\lambda|$ 
is set for the co-annihilation scenario. 
This results in the blank strip for the co-annihilation scenario (red crosses) 
appears at around the region $-10^{-3}<\delta\lambda<0$. 
We note that the loop correction $|\delta\lambda|$ is slightly larger 
for the mixed scenario in the region $\lambda_2 \gsim 1.4$.


\end{document}